\documentclass[amssymb, preprintnumbers, showpacs showkeys,aps,prl,superscriptaddress,twocolumn,nobibnotes]{revtex4-1}

\usepackage{color} 
\usepackage{xcolor} 
\usepackage{hyperref}
\usepackage{slashed}
\usepackage{graphicx}
\usepackage{amsmath}
\usepackage{latexsym}
\usepackage{epstopdf}
\usepackage{amsmath}
\usepackage{enumitem}
\usepackage{amssymb,amsmath}
\usepackage{multirow}
\usepackage{mathtools}
\usepackage{bbm}
\usepackage{bm}
\usepackage{amsfonts}
\usepackage{amssymb}
\usepackage{float}
\usepackage{calligra}
\usepackage{makecell}
\usepackage[normalem]{ulem}
\usepackage[USenglish,american]{babel}
\usepackage{subfigure}

\expandafter\ifx\csname package@font\endcsname\relax\else
 \expandafter\expandafter
 \expandafter\usepackage
 \expandafter\expandafter
 \expandafter{\csname package@font\endcsname}%
\fi
\begin{document}

\title{Coexistence of long- and quasi-long range spatial order in 1D  quantum quasicrystals}

\author{A. Mendoza-Coto}%
\email{alejandro.mendoza@ufsc.br}
\affiliation{Departamento de F\'\i sica, Universidade Federal de Santa Catarina, 88040-900 Florian\'opolis, Brazil}%
\affiliation{Max Planck Institute for the Physics of Complex Systems, Nothnitzerstr. 38, 01187 Dresden, Germany}

\author{M. Grossklags}%
\affiliation{Departamento de F\'\i sica, Universidade Federal de Santa Catarina, 88040-900 Florian\'opolis, Brazil}%

\author{J. Stefaniak}%
\affiliation{Institute for Quantum Electronics, Eidgenossische Technische Hochschule Zurich, Otto-Stern-Weg 1, CH-8093 Zurich, Switzerland}

\author{T. Donner}%
\affiliation{Institute for Quantum Electronics, Eidgenossische Technische Hochschule Zurich, Otto-Stern-Weg 1, CH-8093 Zurich, Switzerland}

\author{F. Piazza}%
\affiliation{Theoretical Physics III, Center for Electronic Correlations and Magnetism,
Institute of Physics, University of Augsburg, 86135 Augsburg, Germany}
\begin{abstract}
    Quasicrystals exhibit long-range positional order without periodicity, arising from multiple incommensurate wave vectors. In self-assembled quasicrystals, the spontaneous breaking of translational invariance gives rise to two Goldstone modes—phonons and phasons—associated with two competing wave vectors. Here, we demonstrate a unique scenario exclusive to quasicrystals: a mechanism that gaps out only one Goldstone mode (associated with one wave vector), while the other remains gapless. In one dimension, this leads to the coexistence of long-range order at the gapped wave vector and quasi-long-range order at the gapless one, due to sustained fluctuations. We show that this phenomenon can be realized using ultracold bosonic atoms in optical cavities.
\end{abstract}
\maketitle
\textit{Introduction.}--Quasicrystals (QCs) feature a spatial pattern that displays long-range positional order in the absence of periodicity. This is due to the presence of multiple incommensurate wave vectors characterizing its structure~\cite{lifshitz2011,savitz2018,jagan2021}. In self-assembled QCs arising through the spontaneous breaking of translational invariance~\cite{gopalakrishnan2013,mendoza2022,grossklags2024,grossklags2025,he2026}, the usual phonons are complemented by an extra set of Goldstone modes, called phasons, that correspond to continuous local rearrangements of the pattern~\cite{lubensky1985,socolar1986}. 
In particular, in one spatial dimension ($1$D), a bi-chromatic QC is characterized by two incommensurate momenta and two Goldstone excitation branches: a phonon and a phason~\cite{socolar1986,ding1993,lifshitz2011}.

The main result of this work is that these features give rise to a scenario unique to quasicrystals: some of the Goldstone modes can persist even in presence of an external breaking of translation invariance. For instance, given a QC pattern with two dominant Fourier components, the Goldstone mode associated with the phase of one Fourier component can acquire a gap while the one related to the phase of the other Fourier mode remains gapless. In $1$D, this leads to the coexistence of long-range positional order at one Fourier component, due to the suppression of fluctuations by the gap, and quasi-long-range order at the other, where the remaining gapless Goldstone mode still supports fluctuations strong enough to destroy true long-range order.

We show that this scenario can be implemented with ultracold bosonic atoms in optical cavities~\cite{mivehvar2019}. The cavity mirrors explicitly break continuous translational invariance by generating an oscillatory, non-decaying atom-atom interaction characterized by a single Fourier component. Additionally, a scanning laser beam~\cite{bonifacio2024} can be designed to endow this interaction with a decaying envelope promoting further modulation with a wave vector that is incommensurate with respect to the cavity one, leading to the competition required for the stabilization of the QC~\cite{mendoza2022,grossklags2024,grossklags2025}.

\begin{figure*}[!t]
    \centering
    \includegraphics[width=\textwidth]{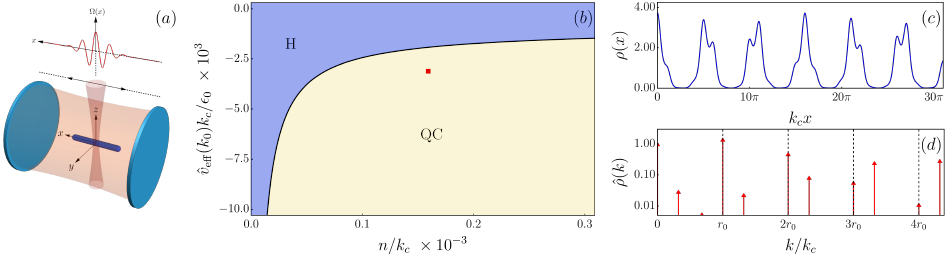}
    \caption{(a) Schematic cavity QED experimental setup for the implementation of laser painted interactions. The elongated BEC is coupled to the standing wave of the cavity and periodically scanned by the structured laser inducing a finite range on demand effective interaction. (b) Ground-state phase diagram of the system for a laser beam profile $\hat{\Omega}(k)$ exciting a finite wave vector $k_{0}$, with $\hat{v}_{\mathrm{eff}}(k_{0})k_{c}/\epsilon_{0}$ and the density $n/k_{c}$ used as running parameters. The red square corresponds to $n/k_{c}=0.159\times10^{-3}$ and $\hat{v}_{\mathrm{eff}}(k_{0})k_{c}/\epsilon_{0}=-3.140\times10^{3}$. (c) and (d) Typical QC density profile in real and momentum space corresponding to the highlighted red square in (b), in (d) $r_{0}$ has been chosen as $\lvert k_{0}-k_{c}\rvert/k_{c}$.}
    \label{fig1}
\end{figure*}

\textit{Model and Method.}--As schematically shown in Fig.~\ref{fig1}(a), we consider a quasi-one-dimensional Bose-Einstein condensate (BEC) in the thermodynamic limit formed by $N$ particles of mass $m$. The condensate extends over a length $L$ along the $x$-direction and is strongly confined by a harmonic trap along the $y$ and $z$ directions. Additionally, the bosonic gas is coupled to a high-finesse cavity containing two parallel highly reflective mirrors placed perpendicular to its axis, while it is periodically scanned by a structured laser-beam along the $x$-direction. For a scanning rate lying between the slow atomic timescales and the faster cavity scales, an effective, time-averaged cavity-mediated interaction is induced, that can be freely designed through the scanning-beam profile. This scheme has been recently proposed and characterized by Bonifacio et al. in Ref.~\cite{bonifacio2024}. Here, we limit ourselves assuming that the cavity-mediated interaction potential of interest has been designed.

Along the transverse $y$ and $z$ directions, the laser profile is taken to be constant, since the illuminated region is assumed to be much larger than the cloud dimensions. Thus, we perform a ``dimensional reduction'' and focus on the one-dimensional problem along the $x$-direction. As a consequence, we obtain the following effective $1$D Hamiltonian (see End Matter section)
\begin{equation}\label{Ham}
    \begin{split}
        \mathcal{H}[\psi,\psi^{\ast}]=&N\int\frac{dx}{L}\bigg\{\frac{\hbar^{2}}{2m}\lvert\partial_{x}\psi(x)\rvert^{2}+\frac{g_{\mathrm{aa}}n}{2}\lvert\psi(x)\rvert^{4}\\
        &+\frac{n}{2}\int dx'\cos\left(\frac{2\pi x}{\lambda_{c}}\right)\cos\left(\frac{2\pi x'}{\lambda_{c}}\right)\\ & \times v_{\mathrm{eff}}(x-x')\lvert\psi(x)\rvert^{2}\lvert\psi(x')\rvert^{2}\bigg\},
    \end{split}
\end{equation}
where $n=N/L$ stands for the linear particle density and $\psi(x)$ satisfies the normalization condition $\int_{0}^{L}dx \lvert\psi(x)\rvert^{2}=L$. The first term of Eq.~\eqref{Ham} accounts for the kinetic energy of the BEC, whereas the second contribution corresponds to the contact interaction with coupling constant $g_{aa}$. Lastly, the non-local contribution corresponds to the laser-painted cavity-mediated interaction, which is a product of two contributions. The cosinusoidal contribution breaks the continuous translational invariance, reducing it to an invariance under a shift by multiples of the cavity wavelength $\lambda_{c}$. Moreover, the kernel $v_{\mathrm{eff}}(x)$ denotes a translational invariant contribution designed via the scanning beam profile (see End Matter for details). Here, we consider the following form
\begin{equation}
    \begin{split}
        v_{\mathrm{eff}}(x)=&-\frac{\Delta s}{\sqrt{2\pi}}\exp\left(-\frac{s^{2}x^{2}}{8}\right)\bigg[\exp\left(-\frac{2k^{2}_{0}}{s^{2}}\right)\\
        &+\cos(k_{0}x)\bigg],
    \end{split}
    \label{pot}
\end{equation}
which features a finite range controlled by $s^{-1}$ and an oscillatory behavior with wave vector $k_{0}$, incommensurate with the cavity wave vector $k_{c}=2\pi/\lambda_{c}$. A negative minimum in $\hat{v}_{\mathrm{eff}}(k)$ promotes an instability at the wave vectors $k_{0}\pm k_{c}$. This mechanism generates the competition between two incommensurate length scales required to stabilize a QC. 

To determine the ground-state of the system, we use a spectral variational method previously employed in the study of supersolids~\cite{prestipino2018,zhang2019,mendoza2021,lima2025}, quantum quasicrystals~\cite{mendoza2022,zampronio2024,grossklags2025} and quasicrystalline Bose glass phases~\cite{grossklags2024}. This method introduces a general ansatz for the wavefunction in the form of a Fourier expansion
\begin{equation}\label{ground_state_wavefunction}
    \psi(x)=\frac{1}{2\Xi}\sum_{\boldsymbol{\mathrm{n}}}c_{\boldsymbol{\mathrm{n}}}\cos(k_{\boldsymbol{\mathrm{n}}}x),
\end{equation}
where $c_{\boldsymbol{\mathrm{n}}}$ and $k_{\boldsymbol{\mathrm{n}}}$ represent the set of Fourier amplitudes and wave vectors of the expansion, respectively. Moreover, the parameter $\Xi=\sqrt{\sum_{\boldsymbol{\mathrm{n}}}c^{2}_{\boldsymbol{\mathrm{n}}}/4}$ ensures the normalization of the wavefunction. The wave vector set of the Fourier expansion is taken as $k_{\boldsymbol{\mathrm{n}}}=n_{1}k_{1}+n_{2}k_{2}$, where $k_{1(2)}$ are treated as variational parameters for the minimization of the energy-per-particle functional $\mathcal{H}[\psi,\psi^{\ast}]/N$. In the present case, the effective interaction couples the Fourier components in $\lvert\psi(x)\rvert^{2}$ to the cavity Fourier components $\exp(\pm ik_{c}x)$. For this reason, the wave vector basis should include $k_{c}$. Hence, without loss of generality, we set $k_{1}=k_{c}$. Moreover, the variational minimization will select a value of $k_{2}$ close to the other characteristic wave vector $k_{0}$. 

\textit{Ground-state phase diagram.}--We use $k_{c}$ as our unit of momentum, as well as $k_{c}^{-1}$, $\epsilon_{0}=\hbar^{2}k_{c}^{2}/m$ and $\hbar/\epsilon_{0}$ as our units of length, energy and time, respectively. The numerical study was performed with the pair potential parameters $s/k_{c}=0.1$, $k_{0}/k_{c}=\gamma-1$, where $\gamma$ stands for the golden ratio, and $g_{\mathrm{aa}}k_{c}/\epsilon_{0}=3.14\times10^{-4}$, a realistic value in current experiments. The phase diagram was determined using $\hat{v}_{\mathrm{eff}}(k_{0})k_{c}/\epsilon_{0}$ and $n/k_{c}$ as running parameters and is presented in Fig.~\ref{fig1}(b). We find a second order transition between the H to QC in which the two main Fourier components of the QC approach zero continuously at the phase boundary and display mean-field scaling, $c_{j}\propto(\hat{v}_{\mathrm{eff}}(k_{0})\vert_{c}-\hat{v}_{\mathrm{eff}}(k_{0}))^{1/2}$, see Supp. Mat.~\cite{supplemental_material}. 
In Fig.~\ref{fig1}(c) and Fig.~\ref{fig1}(d), we present the ground-state density configuration in real and momentum space in a regime with high density contrast. Moreover, we verified that close enough to the QC phase boundary the system develops a finite density background indicating the presence of a finite superfluid fraction, as previously observed in $2$D quantum quasicrystal phases~\cite{mendoza2022,grossklags2024,grossklags2025}.

\textit{Low-energy excitations.}--Now that we have established the existence of a $1$D QC in the present model, we investigate its low-energy excitations properties and hydrodynamic behavior. To this end, we begin by discussing the symmetries of $\mathcal{H}[\psi,\psi^{\ast}]$ under translations of the condensate phase and of the phases associated with the modulated pattern. Therefore, we consider a perturbed wavefunction of the form
\begin{equation}\label{wfunct}
    \tilde{\psi}(x)=\frac{e^{i\theta}}{2\Xi}\sum_{\boldsymbol{\mathrm{n}}}c_{\boldsymbol{\mathrm{n}}}\cos{\big(n_{1}k_{1}(x+\delta_{1})+n_{2}k_{2}(x+\delta_{2})\big)},
\end{equation}
where $\theta$ represents the condensate phase, while $\delta_{1}$ and $\delta_{2}$ are the independent phases associated with the components of the modulated pattern with wave vectors $k_{1}$ and $k_{2}$, respectively. Using the ansatz in Eq.~\eqref{wfunct}, we can compute the change of energy associated with the perturbed wavefunction to find $\mathcal{H}[\tilde{\psi},\tilde{\psi}^{\ast}]-\mathcal{H}[\psi,\psi^{\ast}]\propto(1-\cos(2k_{c}\delta_{1}))$. This shows that the Hamiltonian is invariant under shifts of $\theta$ and $\delta_{2}$, implying in the existence of two gapless Goldstone modes despite the absence of translational invariance. On the other hand, the Hamiltonian is invariant only under discrete shifts of the pattern with the cavity wave vector $k_{1}=k_{c}$, that is, $\delta_{1}\rightarrow\delta_{1}+n\pi/k_{c}$. Therefore, this analysis indicates that the H to QC transition is characterized by a discrete symmetry breaking in the $\delta_{1}$ sector and a BKT-like transition in the $\delta_{2}$ sector~\footnote{The discrete symmetry breaking described here is of the same type as that associated with the Dicke-type superradiant transition in cavity QED systems.}. To better understand the interplay between Goldstone and gapped degrees of freedom, we employ the coherent-state path-integral formalism to construct the Euclidean hydrodynamic action of our system~\cite{mendoza2025}, details on how this construction is done are presented in End Matter section.

\begin{figure}[!t]
    \centering
    \includegraphics[width=0.47\textwidth]{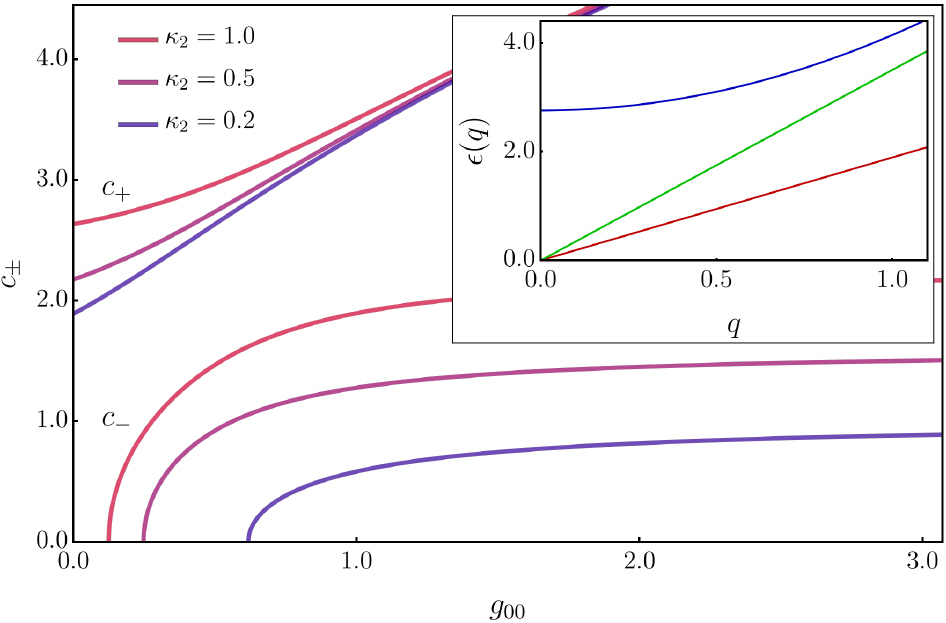}
    \caption{Velocity of the two gapless modes as a function of the hydrodynamic action parameters $g_{00}$ and $\kappa_{2}$, while the others are fixed as follows: $\Gamma=(1+\sqrt{5})/2$, $\alpha_{1}=\alpha_{2}=0.3$, $g_{11}=1$, $g_{22}=3$, $g_{12}=1$, $\kappa_1(q)=1+q^2$ and $n=10$. (see Eq.~\eqref{effective_action} in the End Matter). The inset shows the dispersion relation of the gapped and gapless modes for $\kappa_{2}=1$ and $g_{00}=1$.}
    \label{fig2}
\end{figure}

As shown in the inset of Fig.~\ref{fig2}, the QC phase features three different excitation modes, one gapped mode with quadratic dispersion and two gapless, linearly dispersing modes with velocity $c_{\pm}$. As we shall see next, the gappless modes result from the hybridization between phonon, phason and condensate degrees of freedom.
Indeed, as can be seen from Eq.~\eqref{effective_action} in the End Matter, several terms in the hydrodynamic action couple these degrees of freedom. The usual interaction between phonons and phasons in $1$D quasicrystals is strongly modified in our case by the standing-wave coupling of the cavity. Hybridization is apparent in Fig.~\ref{fig2}, where we plot $c_{\pm}$ as a function of one of the coupling parameters, showing that increasing $\kappa_{2}$ (the stiffness of the phase $\delta_{2}$, see Eq.~\eqref{effective_action}) increases the sound velocity of both gapless modes, confirming the strong hybridization.

The fact that $c_{-}$ vanishes at a critical value is a well known feature produced in our case due to the hybridization of the phonon-phason mode associated with $\delta_{2}(x,\tau)$, but also analogously observed in supersolids and $2$D quasicrystals~\cite{platt2024,poli2024,rakic2024,mendoza2025}. This limit of stability is related to the loss of coherence of the condensate phase and hence of the superfluidity in the system \cite{platt2024,poli2024,rakic2024,mendoza2025}. Regarding the eigenvectors associated with each mode, we conclude that in the long-wavelength limit, the gapped mode satisfies $\hat{\boldsymbol{\mathrm{\eta}}}_{1}(q)=\{O(q^{2}), O(q^{2}), O(1), O(1), O(1), O(1)\}$, whereas the gapless modes satisfy $\hat{\boldsymbol{\mathrm{\eta}}}_{\pm}(q)=\{O(q), O(1), O(q), O(q), 0, O(1)\}$. This allows us to conclude that, at low momentum, the gapless excitations predominantly activate the condensate phase $\theta(x,\tau)$ and the second quasicrystal phase $\delta_{2}(x,\tau)$, in agreement with the analogous situation in supersolids, while leaving the $\delta_{1}(x,\tau)$ phase locked to the phase of the standing wave inside the cavity.

\textit{Properties of the QC order.}--How do the above low-energy excitations property affect the QC order? We focus first on the behavior of the density profile in the presence of a propagating long-wavelength excitation, where we consider that the average particle number over distances larger than the modulation length remains constant. If we assume a general ground-state density profile of the form
$\rho_{0}(x)=\sum_{\boldsymbol{\mathrm{n}}}b_{\boldsymbol{\mathrm{n}}}\cos\big((n_{1}k_{1}+n_{2}k_{2})x\big)$, the presence of a long-wavelength phase deformation $\delta_{2}(x,t)$ leads to the following particle density
\begin{equation}
    \begin{split}
        \rho(x,t)=&\sum_{\boldsymbol{\mathrm{n}}}b_{\boldsymbol{\mathrm{n}}}\cos\big(n_{1}k_{1}x+n_{2}k_{2}(x+\delta_{2}(x,t))\big)\\
        &\times\left(1+\frac{n_{2}k_{2}}{n_{1}k_{1}+n_{2}k_{2}}\partial_{x}\delta_{2}(x,t)\right).
    \end{split}
\end{equation}
This result is a direct generalization of the known behavior of the local density in $1$D supersolids deformed by a long-wavelength lattice displacement $u(x,t)$, where $\rho(x,t)=\rho_{0}(x-u(x,t))(1-\partial_{x}u(x,t))$~\cite{yoo2010,platt2024}. In a quantum QC, each family of Fourier components contributing to a given periodic component behaves as an independent supersolid, consequently, each contribution must satisfy particle-number conservation separately in order for their superposition to do so.

\begin{figure}[!t]
    \centering
    \includegraphics[width=0.44\textwidth]{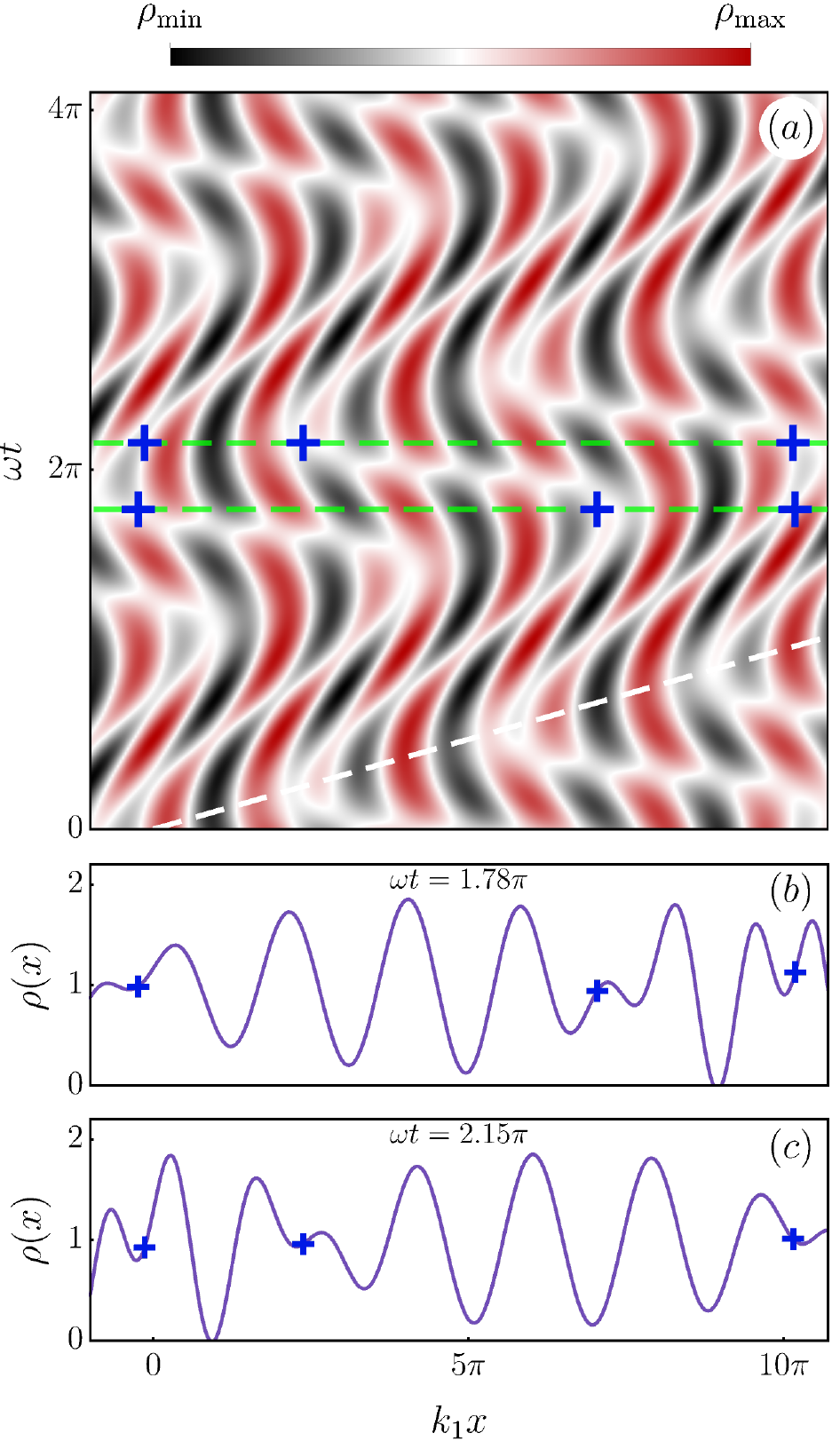}
    \caption{(a) Space-time local density behavior of a $1$D QC phase under the propagation of a long-wavelength excitation in the model considered for gapless excitations. The white dashed line corresponds to the curve $x=c_{s}t$, where $c_{s}$ represents the sound velocity of the excitation mode. The green lines corresponds to two different values of $t$ at which the instantaneous density profiles are presented in (b) and (c). The blue crosses marks the position of the defects in the corresponding modulated patterns.}
    \label{fig3}
\end{figure}

In Fig.~\ref{fig3}(a), we present the space-time behavior of the local density for a simplified quasicrystal whose Fourier expansion contains only two modes with wave vectors $k_{1}$ and $k_{2}$. We consider a single low-energy excitation propagating in the system with $\delta_{2}(x,t)=\mathrm{Re}[\delta_{2,\nu}\exp\left(i(qx-c_{\nu}qt)\right)]$ and $\delta_{1}(x,t)=0$. As can be observed, after a period $T=2\pi/\omega=2\pi/(c_{\nu}q)$, the density pattern returns to the initial aperiodic configuration; nevertheless, one can still identify ripples propagating in the space-time texture, as in the case of (super)solids. As a guide to the eye for these ripples, the white dashed line corresponds to the contour $x=c_{\nu}t$ along which $\delta_{2}(x,t)=\mathrm{const.}$. It is also important to note that the defining feature of our quasicrystal---a locked $\delta_{1}(x,t)$ phase together with a free $\delta_{2}(x,t)$ phase---implies that the propagating excitation generates a space-time texture in which the maxima and minima of the density pattern are not conserved as the system evolves in time. Physically, this is due to the accumulation of a $\delta_{2}(x,t)$ relative phase shift between the periodic components of the quasicrystal as the excitation propagates. In the phonon-phason picture, this behavior follows from enforcing the constraint $u(x,t)-\Gamma w(x,t)=\delta_{1}=0$, which produces a mixed behavior where we have ripples, associated to phonons and discontinuities or defects in the maxima and minima of the spatiotemporal density pattern associated to phasons. This behavior is illustrated in the configurations shown in Figs.~\ref{fig3}(b--c), where the defect positions in the modulated pattern associated with phason dynamics are marked with blue crosses. The dynamics associated to pure phononic $(u(x,t)\neq0$, $w(x,t)=0)$ and pure phasonic $(u(x,t)=0$, $w(x,t)\neq0)$ excitations in the evolution of the density pattern of a system without the $\delta_{1}(x,t)=0$ constraint is presented in the Supp. Mat.~\cite{supplemental_material}. For the sake of comparison, we consider in each case excitation fields $u(x,t)$ or $w(x,t)$ with the same amplitude, frequency and velocity with respect to the one of $\delta_{2}(x,t)$ in Fig.~\ref{fig3}.

Lastly, we analyze the role of spatial $\delta_{1}(x,t)$ and $\delta_{2}(x,t)$ phase fluctuations in the density pattern. The phase correlations can be directly calculated in our case by inverting the interaction matrix $\boldsymbol{\mathrm{M}}(q,\omega)$. Indeed, we have $\langle\hat{\delta}_{1}(q,\omega)\hat{\delta}_{1}(q',\omega')\rangle=(2\pi)^{2}\delta(q+q')\delta(\omega+\omega')C_{1}(q,\omega)$, where $C_{1}(q,\omega)=\boldsymbol{\mathrm{M}}(q,\omega)_{5,5}^{-1}(q,\omega)$ and $\langle\hat{\delta}_{2}(q,\omega)\hat{\delta}_{2}(q',\omega')\rangle=(2\pi)^{2}\delta(q+q')\delta(\omega+\omega')C_{2}(q,\omega)$, in which $C_{2}(q,\omega)=\boldsymbol{\mathrm{M}}(q,\omega)_{6,6}^{-1}(q,\omega)$. Despite the fact that the elements of $\boldsymbol{\mathrm{M}}(q,\omega)^{-1}$ have a convoluted form, it is possible to conclude that in the long-wavelength limit $(q\rightarrow0)$, we have $C_{1}(q,\omega)=(\kappa_{1}+\gamma_{11} \omega^{2}+\gamma_{12}q^{2})^{-1}$ and $C_{2}(q,\omega)=(\gamma_{21}\omega^{2}+\gamma_{22}q^{2})^{-1}$, where the $\gamma$ parameters are convoluted rational functions of the couplings of the interaction matrix, see Supp. Mat.~\cite{supplemental_material}. As a consequence, neglecting $\delta_{1}-\delta_{2}$ correlations, we can conclude that
\begin{equation}\label{average_dens}
    \begin{split}
        \langle\rho(x)\rangle=&\sum_{\boldsymbol{\mathrm{n}}}b_{\boldsymbol{\mathrm{n}}}\cos\big((k_{1}n_{1}+k_{2}n_{2})x\big)\\
        &\times\exp\big(-\frac{(k_{1}n_{2})^{2}}{2}\langle\delta^{2}_{1}\rangle-\frac{(k_{2}n_{2})^{2}}{2}\langle\delta^{2}_{2}\rangle\big),
    \end{split}
\end{equation}
where the variances are given by $\langle\delta^{2}_{1}\rangle=\ln(2\pi\sqrt{\gamma_{11}/\kappa}/a)/(2\pi\sqrt{\gamma_{11}\gamma_{12}})$ and $\langle\delta^{2}_{2}\rangle=\ln(L/a)/(2\pi\sqrt{\gamma_{21}\gamma_{22}})$, with $L$ and $a$ corresponding to the linear dimension of the system and the short distance cut-off, respectively. As can be observed in Eq.~\eqref{average_dens}, the Fourier components of the density pattern with momentum $n_{1}k_{1}$ ``renormalizes'' the expression by a finite factor $(2\pi/a\sqrt{\gamma_{11}/\kappa})^{-k^{2}_{1}n^{2}_{1}/(4\pi\sqrt{\gamma_{11}\gamma_{12}})}$ in the $L\rightarrow\infty$ limit, whereas those with momenta $n_{2}k_{2}$ reescale by a factor $(a/L)^{k^{2}_{2}n^{2}_{2}/(4\pi\sqrt{\gamma_{21}\gamma_{22}})}$. These results show that the component of the modulated pattern with momentum proportional to $k_{1}$ displays long-range positional order, while that with momenta proportional to $k_{2}$, decaying with the characteristic power law behavior of the system size $(L/a)$, has positional quasi-long-range order.

\textit{Conclusions.}--We have shown that a one-dimensional quantum quasicrystal can realize a form of spatial order in which different Fourier components of the same quasiperiodic structure display distinct infrared behavior. In the cavity-based implementation considered here, the standing-wave contribution explicitly breaks continuous translation invariance only with respect to the cavity wave vector. As a consequence, one linear combination of the quasicrystalline phases becomes pinned and acquires a gap, while the orthogonal combination remains gapless. This selective gapping of Goldstone modes leads to the coexistence of true long-range order at the pinned wave vector family and quasi-long-range order at the unpinned one.

This result provides a quasicrystalline counterpart to several well-known phenomena in low-dimensional systems, while differing from each of them in an essential way. In one-dimensional quantum crystals, superfluids, and supersolids, gapless phase or displacement fluctuations destroy true long-range positional order, leaving only algebraic correlations. In contrast, pinned density waves, commensurate charge-density waves, and Frenkel--Kontorova-type systems acquire true long-range order because the sliding mode is locked by an external potential or by commensurability. The present system combines both behaviors within a single quasiperiodic phase: one component behaves as a pinned density wave, whereas the other retains the algebraic order characteristic of a fluctuating one-dimensional quantum system.\\

\textit{Acknowledgments}--A.M.C. acknowledges financial support from the Brazilian National Council for Scientific and Technological Development (CNPq), under Grant No. 311122/2026-4, and thanks the Max Planck Institute for the Physics of Complex Systems (MPIPKS) for its financial support and hospitality. F.P. acknowledges funding from the Munich Quantum Valley within the Hightech Agenda Bayern Plus supported by the State Ministry of Science and the Arts. This research was supported by the Swiss National Science Foundation (SNSF) project numbers 217124, 221538, 223274, and the Swiss State Secretariat for Education, Research and Innovation (SERI) under grant number MB22.00090. This project is funded within the QuantERA II Programme, which has received funding from the EU's Horizon 2020 research and innovation programme under Grant Agreement No. 101017733.

\bibliography{ref.bib} 

@article{mendoza2022,
    title = {Exploring quantum quasicrystal patterns: A variational study},
    author = {Mendoza-Coto, A. and Turcati, R. and Zampronio, V. and D\'{\i}az-M\'endez, R. and Macr\`{\i}, T. and Cinti, F.},
    journal = {Phys. Rev. B},
    volume = {105},
    issue = {13},
    pages = {134521},
    numpages = {11},
    year = {2022},
    month = {Apr},
    publisher = {American Physical Society},
    doi = {10.1103/PhysRevB.105.134521},
    url = {https://link.aps.org/doi/10.1103/PhysRevB.105.134521}
}

@article{mivehvar2019,
    title = {Emergent Quasicrystalline Symmetry in Light-Induced Quantum Phase Transitions},
    author = {Mivehvar, Farokh and Ritsch, Helmut and Piazza, Francesco},
    journal = {Phys. Rev. Lett.},
    volume = {123},
    issue = {21},
    pages = {210604},
    numpages = {7},
    year = {2019},
    month = {Nov},
    publisher = {American Physical Society},
    doi = {10.1103/PhysRevLett.123.210604},
    url = {https://link.aps.org/doi/10.1103/PhysRevLett.123.210604}
}

@article{gopalakrishnan2013,
    title = {Quantum Quasicrystals of Spin-Orbit-Coupled Dipolar Bosons},
    author = {Gopalakrishnan, Sarang and Martin, Ivar and Demler, Eugene A.},
    journal = {Phys. Rev. Lett.},
    volume = {111},
    issue = {18},
    pages = {185304},
    numpages = {5},
    year = {2013},
    month = {Oct},
    publisher = {American Physical Society},
    doi = {10.1103/PhysRevLett.111.185304},
    url = {https://link.aps.org/doi/10.1103/PhysRevLett.111.185304}
}

@article{grossklags2025,
    title = {Engineering interaction potentials for stabilizing quantum quasicrystal phases},
    author = {Grossklags, Matheus and Lima, Daniel and Zampronio, Vinicius and Cinti, Fabio and Mendoza-Coto, Alejandro},
    journal = {Phys. Rev. B},
    volume = {112},
    issue = {22},
    pages = {224107},
    numpages = {15},
    year = {2025},
    month = {Dec},
    publisher = {American Physical Society},
    doi = {10.1103/vnt6-9g1l},
    url = {https://link.aps.org/doi/10.1103/vnt6-9g1l}
}

@article{prestipino2018,
    title = {Freezing of soft-core bosons at zero temperature: A variational theory},
    author = {Prestipino, Santi and Sergi, Alessandro and Bruno, Ezio},
    journal = {Phys. Rev. B},
    volume = {98},
    issue = {10},
    pages = {104104},
    numpages = {16},
    year = {2018},
    month = {Sep},
    publisher = {American Physical Society},
    doi = {10.1103/PhysRevB.98.104104},
    url = {https://link.aps.org/doi/10.1103/PhysRevB.98.104104}
}

@article{bonifacio2024,
    title = {Laser-Painted Cavity-Mediated Interactions in a Quantum Gas},
    author = {Bonifacio, Mariano and Piazza, Francesco and Donner, Tobias},
    journal = {PRX Quantum},
    volume = {5},
    issue = {4},
    pages = {040332},
    numpages = {11},
    year = {2024},
    month = {Dec},
    publisher = {American Physical Society},
    doi = {10.1103/PRXQuantum.5.040332},
    url = {https://link.aps.org/doi/10.1103/PRXQuantum.5.040332}
}

@misc{he2026,
    title={Self-organized quasicrystals and their excitations in dipolar Bose-Einstein condensates via optical feedback}, 
    author={Liang-Jun He and Fabian Maucher and Yong-Chang Zhang},
    year={2026},
    eprint={2607.10141},
    archivePrefix={arXiv},
    primaryClass={cond-mat.quant-gas},
    url={https://arxiv.org/abs/2607.10141}, 
}

@article{grossklags2024,
    title = {Self-induced Bose glass phase in quantum quasicrystals},
    journal = {Results in Physics},
    volume = {65},
    pages = {107991},
    year = {2024},
    issn = {2211-3797},
    doi = {https://doi.org/10.1016/j.rinp.2024.107991},
    url = {https://www.sciencedirect.com/science/article/pii/S2211379724006764},
    author = {M. Grossklags and M. Ciardi and V. Zampronio and F. Cinti and A. Mendoza-Coto}
}

@article{zhang2019,
    title = {Supersolidity around a Critical Point in Dipolar Bose-Einstein Condensates},
    author = {Zhang, Yong-Chang and Maucher, Fabian and Pohl, Thomas},
    journal = {Phys. Rev. Lett.},
    volume = {123},
    issue = {1},
    pages = {015301},
    numpages = {6},
    year = {2019},
    month = {Jul},
    publisher = {American Physical Society},
    doi = {10.1103/PhysRevLett.123.015301},
    url = {https://link.aps.org/doi/10.1103/PhysRevLett.123.015301}
}

@article{lima2025,
    title = {Supersolid dipolar phases in planar geometry: Effects of tilted polarization},
    author = {Lima, Daniel and Grossklags, Matheus and Zampronio, Vinicius and Cinti, Fabio and Mendoza-Coto, Alejandro},
    journal = {Phys. Rev. A},
    volume = {111},
    issue = {6},
    pages = {063311},
    numpages = {11},
    year = {2025},
    month = {Jun},
    publisher = {American Physical Society},
    doi = {10.1103/mlfn-114m},
    url = {https://link.aps.org/doi/10.1103/mlfn-114m}
}

@book{stoof2009,
    title={Ultracold quantum fields},
    author={Stoof, Henk TC and Gubbels, Koos B and Dickerscheid, Dennis BM},
    year={2009},
    publisher={Springer}
}

@article{mendoza2025,
    title = {Low-Energy Excitations in Bosonic Quantum Quasicrystals},
    author = {Mendoza-Coto, A. and Bonifacio, M. and Piazza, F.},
    journal = {Phys. Rev. Lett.},
    volume = {134},
    issue = {13},
    pages = {136003},
    numpages = {7},
    year = {2025},
    month = {Apr},
    publisher = {American Physical Society},
    doi = {10.1103/PhysRevLett.134.136003},
    url = {https://link.aps.org/doi/10.1103/PhysRevLett.134.136003}
}

@article{mendoza2021,
    title = {Ground-state phase diagram of ultrasoft bosons},
    author = {Mendoza-Coto, Alejandro and Caetano, Diogo de Souza and D\'{\i}az-M\'endez, Rogelio},
    journal = {Phys. Rev. A},
    volume = {104},
    issue = {1},
    pages = {013301},
    numpages = {7},
    year = {2021},
    month = {Jul},
    publisher = {American Physical Society},
    doi = {10.1103/PhysRevA.104.013301},
    url = {https://link.aps.org/doi/10.1103/PhysRevA.104.013301}
}

@article{zampronio2024,
    title = {Exploring Quantum Phases of Dipolar Gases through Quasicrystalline Confinement},
    author = {Zampronio, Vinicius and Mendoza-Coto, Alejandro and Macr\`{\i}, Tommaso and Cinti, Fabio},
    journal = {Phys. Rev. Lett.},
    volume = {133},
    issue = {19},
    pages = {196001},
    numpages = {7},
    year = {2024},
    month = {Nov},
    publisher = {American Physical Society},
    doi = {10.1103/PhysRevLett.133.196001},
    url = {https://link.aps.org/doi/10.1103/PhysRevLett.133.196001}
}

@article{pretko2018,
    title = {Symmetry-Enriched Fracton Phases from Supersolid Duality},
    author = {Pretko, Michael and Radzihovsky, Leo},
    journal = {Phys. Rev. Lett.},
    volume = {121},
    issue = {23},
    pages = {235301},
    numpages = {7},
    year = {2018},
    month = {Dec},
    publisher = {American Physical Society},
    doi = {10.1103/PhysRevLett.121.235301},
    url = {https://link.aps.org/doi/10.1103/PhysRevLett.121.235301}
}

@article{lifshitz2011,
    author = {Lifshitz, Ron},
    title = {Symmetry Breaking and Order in the Age of Quasicrystals},
    journal = {Israel Journal of Chemistry},
    volume = {51},
    number = {11-12},
    pages = {1156-1167},
    doi = {https://doi.org/10.1002/ijch.201100156},
    url = {https://onlinelibrary.wiley.com/doi/abs/10.1002/ijch.201100156},
    year = {2011}
}

@article{savitz2018,
    title = {Multiple-scale structures: from Faraday waves to soft-matter quasicrystals},
    journal = {IUCrJ},
    volume = {5},
    pages = {247-268},
    year = {2018},
    issn = {2052-2525},
    doi = {https://doi.org/10.1107/S2052252518001161},
    url = {https://www.sciencedirect.com/science/article/pii/S2052252522004717},
    author = {Samuel Savitz and Mehrtash Babadi and Ron Lifshitz},
}

@article{jagan2021,
    title = {The Fibonacci quasicrystal: Case study of hidden dimensions and multifractality},
    author = {Jagannathan, Anuradha},
    journal = {Rev. Mod. Phys.},
    volume = {93},
    issue = {4},
    pages = {045001},
    numpages = {37},
    year = {2021},
    month = {Nov},
    publisher = {American Physical Society},
    doi = {10.1103/RevModPhys.93.045001},
    url = {https://link.aps.org/doi/10.1103/RevModPhys.93.045001}
}

@article{socolar1986,
    title = {Phonons, phasons, and dislocations in quasicrystals},
    author = {Socolar, Joshua E. S. and Lubensky, T. C. and Steinhardt, Paul J.},
    journal = {Phys. Rev. B},
    volume = {34},
    issue = {5},
    pages = {3345--3360},
    numpages = {0},
    year = {1986},
    month = {Sep},
    publisher = {American Physical Society},
    doi = {10.1103/PhysRevB.34.3345},
    url = {https://link.aps.org/doi/10.1103/PhysRevB.34.3345}
}

@article{lubensky1985,
    title = {Hydrodynamics of icosahedral quasicrystals},
    author = {Lubensky, T. C. and Ramaswamy, Sriram and Toner, John},
    journal = {Phys. Rev. B},
    volume = {32},
    issue = {11},
    pages = {7444--7452},
    numpages = {0},
    year = {1985},
    month = {Dec},
    publisher = {American Physical Society},
    doi = {10.1103/PhysRevB.32.7444},
    url = {https://link.aps.org/doi/10.1103/PhysRevB.32.7444}
}

@article{ding1993,
    title = {Generalized elasticity theory of quasicrystals},
    author = {Ding, Di-hua and Yang, Wenge and Hu, Chengzheng and Wang, Renhui},
    journal = {Phys. Rev. B},
    volume = {48},
    issue = {10},
    pages = {7003--7010},
    numpages = {0},
    year = {1993},
    month = {Sep},
    publisher = {American Physical Society},
    doi = {10.1103/PhysRevB.48.7003},
    url = {https://link.aps.org/doi/10.1103/PhysRevB.48.7003}
}

@article{platt2024,
    title = {Sound waves and fluctuations in one-dimensional supersolids},
    author = {Platt, L. M. and Baillie, D. and Blakie, P. B.},
    journal = {Phys. Rev. A},
    volume = {110},
    issue = {2},
    pages = {023320},
    numpages = {14},
    year = {2024},
    month = {Aug},
    publisher = {American Physical Society},
    doi = {10.1103/PhysRevA.110.023320},
    url = {https://link.aps.org/doi/10.1103/PhysRevA.110.023320}
}

@article{poli2024,
    title = {Excitations of a two-dimensional supersolid},
    author = {Poli, E. and Baillie, D. and Ferlaino, F. and Blakie, P. B.},
    journal = {Phys. Rev. A},
    volume = {110},
    issue = {5},
    pages = {053301},
    numpages = {10},
    year = {2024},
    month = {Nov},
    publisher = {American Physical Society},
    doi = {10.1103/PhysRevA.110.053301},
    url = {https://link.aps.org/doi/10.1103/PhysRevA.110.053301}
}

@article{yoo2010,
    title = {Hydrodynamic theory of supersolids: Variational principle, effective Lagrangian, and density-density correlation function},
    author = {Yoo, C.-D. and Dorsey, Alan T.},
    journal = {Phys. Rev. B},
    volume = {81},
    issue = {13},
    pages = {134518},
    numpages = {12},
    year = {2010},
    month = {Apr},
    publisher = {American Physical Society},
    doi = {10.1103/PhysRevB.81.134518},
    url = {https://link.aps.org/doi/10.1103/PhysRevB.81.134518}
}

@article{rakic2024,
    title = {Elastic properties and thermodynamic anomalies of supersolids},
    author = {Rakic, Milan and Ho, Andrew F. and Lee, Derek K. K.},
    journal = {Phys. Rev. Res.},
    volume = {6},
    issue = {4},
    pages = {043040},
    numpages = {19},
    year = {2024},
    month = {Oct},
    publisher = {American Physical Society},
    doi = {10.1103/PhysRevResearch.6.043040},
    url = {https://link.aps.org/doi/10.1103/PhysRevResearch.6.043040}
}

@misc{supplemental_material,
    title = {{S}upplemental {M}aterials to {C}oexistence of long- and quasi-long range spatial order in 1{D}  quantum quasicrystals.},
    author = {A. Mendoza-Coto and M. Grossklags and J. Stefaniak and T. Donner and F. Piazza},
    year = {2026}
}
\onecolumngrid
\section*{End Matter}
\twocolumngrid
\textit{Dimensional reduction.}--In this work, we consider a quasi-one-dimensional BEC composed of $N$ particles of mass $m$, strongly confined along the $y$ and $z$ directions and free to move along the $x$-direction. The BEC is coupled to a quantum cavity containing two parallel reflective mirrors placed perpendicular to the axis of the system. In this configuration, the sample is scanned periodically with a structured laser along the $x$-direction which induces an effective attractive cavity mediated interaction. Considering that the laser profile is constant along the transversal directions, the total energy per particle of the condensate can be expressed as
\begin{equation}\label{three_dimensional_energy_functional}
        \begin{split}
            &\frac{E[\Psi(\boldsymbol{r})]}{N}=\int d^{3}r\bigg\{\frac{\hbar^{2}}{2m}\lvert\boldsymbol{\nabla}\Psi(\boldsymbol{r})\rvert^{2}+U(y,z)\lvert\Psi(\boldsymbol{r})\rvert^{2}\\
            &+\frac{g_{\mathrm{aa}}N}{2}\lvert\Psi(\boldsymbol{r})\rvert^{4}+\frac{N}{2}\int d^{3}r'V_{\mathrm{eff}}(x,x')\lvert\Psi(\boldsymbol{r})\rvert^{2}\lvert\Psi(\boldsymbol{r}')\rvert^{2}\bigg\},
        \end{split}
\end{equation}
where $\Psi(\boldsymbol{r})$ represents the BEC wavefunction satisfying the normalization condition $\int d^{3}r\lvert\Psi(\boldsymbol{r})\rvert=1$. The first and second term of Eq.~\eqref{three_dimensional_energy_functional} accounts for the kinetic energy of particles and their potential energy due to the confining potential $U(y,z)=\frac{1}{2}m(\omega^{2}_{y}y^{2}+\omega^{2}_{z}z^{2})$, respectively, where $\omega_{y}$ and $\omega_{z}$ denotes the corresponding trap frequencies along the $y$ and $z$ directions. The third contribution stands for the mean-field approximation of the zero-range collision interaction between particles, in which $g_\mathrm{aa}$ represents the contact interaction energy coupling. The last contribution in Eq.~\eqref{three_dimensional_energy_functional} corresponds to the effective cavity-induced interaction $V_{\mathrm{eff}}(x,x')$, whose form is given by
\begin{equation}\label{V_eff_definition}
    V_{\mathrm{eff}}(x,x')=\cos\left(\frac{2\pi x}{\lambda_{c}}\right)\cos\left(\frac{2\pi x'}{\lambda_{c}}\right)v_\mathrm{eff}(x-x'),
\end{equation}
where $\lambda_{c}$ denotes the cavity wavelength of the system. The translationally invariant part of the effective interaction, $v_{\mathrm{eff}}(x)$, is determined by the profile of the scanning laser mode $\hat{\Omega}(k)$ according to the relation~\cite{bonifacio2024}
\begin{equation}
    v_\mathrm{eff}(x)=-\frac{1}{2\pi}\int dk\exp(ikx)\Delta\hat{\Omega}^{2}(k).
\end{equation}
In our case, we consider a laser mode momentum profile peaked around a finite wave vector $k_{0}$, with width $s$ and amplitude $\omega$, captured by the Gaussian ansatz
\begin{equation}\label{laser_mode_profile}
    \hat{\Omega}(k)=\omega\left(\exp\left(-\frac{(k-k_{0})^{2}}{s^{2}}\right)+\exp\left(-\frac{(k+k_{0})^{2}}{s^{2}}\right)\right).
\end{equation}
Finally, $\Delta$ is a positive parameter that depends on the microscopic details of the cavity-mediated interaction.

We now consider the regime where the transverse trapping frequencies are sufficiently strong to produce a simple single-cloud pattern. In this regime, we perform a ``dimensional reduction'' and concentrate on the more complex one-dimensional self-assembly problem occurring along the $x$-direction of the system. To this end, we propose a three-dimensional wavefunction of the form $\Psi(\boldsymbol{r})=\psi_\perp(y,z)\psi(x)/\sqrt{L}$, where
\begin{equation}\label{three_dimensional_wavefunction}
    \psi_\perp(y,z)=\sqrt{\frac{2}{\pi\sigma_{y}\sigma_{z}}}\exp\left(-\frac{y^{2}}{\sigma^{2}_{y}}-\frac{z^{2}}{\sigma^{2}_{z}}\right),
\end{equation}
with $L$ representing the length of the system along the $x$-direction, and $\sigma_{y}$ and $\sigma_{z}$ denoting the characteristic widths along the $y$ and $z$ directions, respectively. As a consequence, the normalization condition for $\Psi(\boldsymbol{r})$ immediately translates into $\int dx\lvert\psi(x)\rvert^{2}=L$. After integrating over the transverse directions in Eq~\eqref{three_dimensional_energy_functional}, using the ansatz in Eq.~\eqref{three_dimensional_wavefunction}, we obtain the energy per particle of our effective one-dimensional system
\begin{equation}\label{one_dimensional_energy_functional}
    \begin{split}
        \frac{E[\psi(x)]}{N}=&\int\frac{dx}{L}\bigg\{\frac{\hbar^{2}}{2m}\lvert\partial_{x}\psi(x)\rvert^{2}+\frac{\hbar^{2}}{2m}\left(\frac{1}{2\sigma_{y}^{2}}+\frac{1}{2\sigma_{z}^{2}}\right)\\
        &+\frac{1}{8}\left(\omega_{y}^{2}\sigma_{y}^{2}+\omega_{z}^{2}\sigma_{z}^{2}\right)+\frac{g_{\mathrm{aa}}n}{2\pi\sigma_{y}\sigma_{z}}\lvert\psi(x)\rvert^{4}\\
        &+\frac{n}{2}\int dx'V_{\mathrm{eff}}(x,x')\lvert\psi(x)\rvert^{2}\lvert\psi(x')\rvert^{2}\bigg\},
    \end{split}
\end{equation}
where $n=N/L$ represents the linear density of particles along the $x$-direction. This expression can be further simplified by considering that in the strong trapping regime $\sigma_{y(z)}=(2\hbar^{2}/(m\omega_{y(z)}^{2}))^{1/4}$ to leading-order in $\omega_{y(z)}$. 
As a consequence, the ``renormalized'' contact interaction coupling, $\tilde{g}_{\mathrm{aa}}=g_{\mathrm{aa}}/(\pi\sigma_{y}\sigma_{z})$, reduces to $(2\hbar^2)^{1/2}g_{\mathrm{aa}}/(\pi(m^2\omega_{y}\omega_{z})^{1/4})$, a control parameter independent of the system density $n$. Therefore, up to an additive background energy due to confinement along the transversal directions, the Hamiltonian of our effective $1$D model can be written as
\begin{equation}
    \begin{split}
        \mathcal{H}[\psi,\psi^{*}]=&N\int\frac{dx}{L}\bigg\{\frac{\hbar^{2}}{2m}\lvert\partial_{x}\psi(x)\rvert^{2}+\frac{\tilde{g}_{\mathrm{aa}}n}{2}\lvert\psi(x)\rvert^{4}\\
        &+\frac{n}{2}\int dx'V_{\mathrm{eff}}(x,x')\lvert\psi(x)\rvert^{2}\lvert\psi(x')\rvert^{2}\bigg\},
    \end{split}
\end{equation}
which corresponds to Eq.~\eqref{Ham} of the main text.

\begin{figure}[!h]
    \centering
    \includegraphics[width=0.44\textwidth]{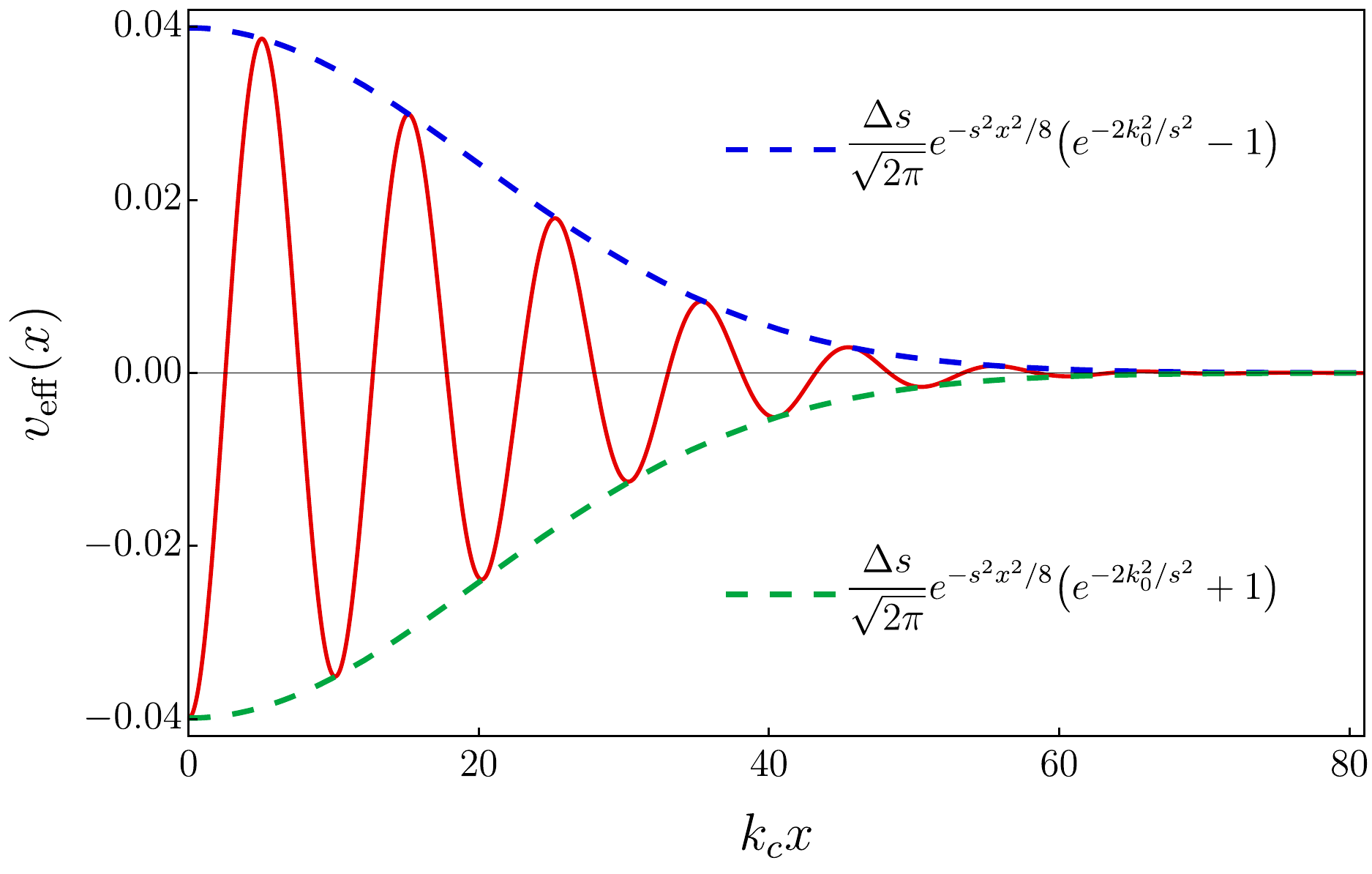}
    \caption{Real-space behavior of the cavity-mediated interaction $v_{\mathrm{eff}}(x)$, shown as a solid red line. Dashed blue and green lines correspond to the upper and lower envelopes of the potential, respectively.}
    \label{fig4}
\end{figure}

\textit{Experimental beam design.}--To favor the stabilization of the QC phase, we consider a laser mode profile $\hat{\Omega}(k)$ that predominantly excites a finite wave vector $k_{0}$, chosen to be incommensurate with the cavity momentum $k_{c}=2\pi/\lambda_{c}$. The Gaussian ansatz introduced in Eq.~\eqref{laser_mode_profile} provides enough flexibility to control both the position of the dominant unstable mode, through $k_{0}$, and the width of the unstable region in momentum space, through the parameter $s$. In the calculations presented here, we set $s/k_{c}=0.1$, which makes this region sufficiently narrow so that the instability is essentially associated with the wave vector $k_{0}$. Taking the inverse Fourier transform of $\hat{\Omega}(k)^{2}$, we reach the real-space interaction $v_{\mathrm{eff}}(x)$ given in Eq.~\eqref{pot}. Moreover, Fig.~\ref{fig4} shows the resulting real-space form of $v_{\mathrm{eff}}(x)$ for $\Delta=1.0$, $s/k_{c}=0.1$, and $k_{0}/k_{c}=\gamma-1$. The solid red line corresponds to the cavity-mediated interaction, while the blue and green dashed lines indicate its upper and lower envelopes, respectively. These envelopes highlight the Gaussian decay of the nonlocal interaction, with a characteristic length scale proportional to $2/s$.

\textit{Low energy effective action.}--A field ansatz compatible with the symmetries of the model considered in the present work, which includes the required long-wavelength fluctuations of the phases and their corresponding conjugate densities for the QC, assumes the form
\begin{equation}\label{wavefunction_fluctuations}
    \begin{split}
        &\phi(x,\tau)=\frac{\sqrt{n}}{2\Xi}\sum\limits_{\boldsymbol{\mathrm{n}}}c_{\boldsymbol{\mathrm{n}}}\bigg(1+\frac{\delta n(x,\tau)}{n}+\boldsymbol{\mathrm{Q}}_{\boldsymbol{\mathrm{n}}}\cdot\boldsymbol{\mathrm{Z}}^{-1}\\
        &\cdot\boldsymbol{\mathrm{\Pi}}(x,\tau)^{\mathrm{T}}\bigg)^{1/2}\exp\bigg(i\boldsymbol{\mathrm{Q}}_{\boldsymbol{\mathrm{n}}}\cdot\big(\boldsymbol{\mathrm{X}}^{\mathrm{T}}+\boldsymbol{\mathrm{U}}(x,\tau)^{\mathrm{T}}\big)+i\theta(x,\tau)\bigg),
    \end{split}
\end{equation}
where $\boldsymbol{\mathrm{Q}}_{\boldsymbol{\mathrm{n}}}=\{n_{1}k_{1}+n_{2}k_{2},-n_{1}k_{2}+n_{2}k_{1}\}$, $\boldsymbol{\mathrm{\Pi}}(x,\tau)=\{\Pi(x,\tau),\Pi_{\perp}(x,\tau)\}$, $\boldsymbol{\mathrm{X}}=\{x,0\}$, and $\boldsymbol{\mathrm{U}}(x,\tau)=\{u(x,\tau),w(x,\tau)\}$. Moreover, the fields $\theta(x,\tau)$, $u(x,\tau)$ and $w(x,\tau)$ represent the complex phase, phonon, and phason deformation fields, whereas $\delta n(x,\tau)$, $\Pi(x,\tau)$ and $\Pi_{\perp}(x,\tau)$ denote the corresponding conjugate density fields. Finally, the $2\times2$ matrix $\boldsymbol{\mathrm{Z}}$ is chosen so that the density vector field $\boldsymbol{\mathrm{\Pi}}(x,\tau)$ and the vector phase field $\boldsymbol{\mathrm{U}}(x,\tau)$ behave as canonically conjugate variables in the full hydrodynamic action; see Supp. Mat.~\cite{supplemental_material} for details. 

Under the assumption that all fluctuation fields introduced above vary slowly in space and imaginary time, we can derive the effective hydrodynamic action by inserting the wavefunction ansatz into the microscopic action and expanding up to quadratic order. This procedure leads to the following effective action
\begin{equation}\label{effective_action}
    \begin{split}
        \delta\mathcal{S}=&\int d\tau dx\bigg\{i\delta n(x,\tau)\partial_{\tau}\theta(x,\tau)+i\Pi(x,\tau)\partial_{\tau}u(x,\tau)\\
        &+i\Pi_{\perp}(x,\tau)\partial_{\tau}w(x,\tau)+\frac{1}{2}\bigg[n\big(\partial_{x}\theta(x,\tau)\big)^{2}\\
        &+g_{00}\big(\delta n(x,\tau)\big)^{2}+g_{11}\Pi(x,\tau)^{2}+g_{22}\Pi_{\perp}(x,\tau)^{2}\\
        &+2g_{12}\Pi(x,\tau)\Pi_{\perp}(x,\tau)+2\Pi(x,\tau)\partial_{x}\theta(x,\tau)\\
        &+2\alpha_{1}\delta n(x,\tau)\partial_{x}u(x,\tau)+\kappa_{1}\delta_{1}(x,\tau)^{2}\\
        &+2\alpha_{2}\delta n(x,\tau)\partial_{x}w(x,\tau)+\kappa_{2}\big(\partial_{x}\delta_{2}(x,\tau)\big)^{2}\bigg]\bigg\},
    \end{split}
\end{equation}
where the auxiliary phase fields $\delta_{1}(x,\tau)$ and $\delta_{2}(x,\tau)$ are defined in terms of the phonon and phason fields as $\delta_{1}(x,\tau)=u(x,\tau)-\Gamma w(x,\tau)$ and $\delta_{2}(x,\tau)=u(x,\tau)+w(x,\tau)/\Gamma$, with $\Gamma=k_{2}/k_{1}$. Moreover, the effective couplings in the action, $g_{i}$, $\alpha_{i}$ and $\kappa_{i}$, can be estimated from the action value calculated up to quadratic order in the fluctuation fields, for details see Supp. Mat.~\cite{supplemental_material}. As expected from the symmetry analysis, when the phonon-phason elastic energy is expressed in terms of the auxiliary fields $\delta_{1}(x,\tau)$ and $\delta_{2}(x,\tau)$, we find that the structure is pinned with respect to $\delta_{1}(x,\tau)$ and remains invariant under constant shifts of $\delta_{2}(x,\tau)$.

Once $u(x,\tau)$ and $w(x,\tau)$ are recast in terms of $\delta_{1}(x,\tau)$ and $\delta_{2}(x,\tau)$, the effective action in Eq.~\eqref{effective_action} can be written in the general form $\delta\mathcal{S}=\frac{1}{2}\int\frac{d\omega}{(2\pi)}\frac{dq}{(2\pi)}\hat{\boldsymbol{\mathrm{n}}}(q,\omega)^{\dagger}\boldsymbol{\mathrm{M}}(q,\omega)\hat{\boldsymbol{\mathrm{n}}}(q,\omega)$, where $\hat{\boldsymbol{\mathrm{n}}}(q,\omega)=(\delta\hat{n}(q,\omega),
\hat{\theta}(q,\omega),\hat{\Pi}(q,\omega),\hat{\Pi}_{\perp}(q,\omega),\hat{\delta}_{1}(q,\omega),\hat{\delta}_{2}(q,\omega))$. Additionally, from the determination of the interaction matrix $\boldsymbol{\mathrm{M}}(q,\omega)$, the excitation-energy spectrum can be calculated by solving the equation $\det\big[\boldsymbol{\mathrm{M}}(q,i\omega(q))\big]=0$~\cite{stoof2009}.


\widetext
\begin{center}
\textbf{\large Supplemental Material: Coexistence of long- and quasi-long range spatial order in 1D  quantum quasicrystals }
\end{center}

\setcounter{equation}{0}
\setcounter{figure}{0}
\setcounter{table}{0}
\setcounter{page}{1}
\makeatletter
\renewcommand{\theequation}{S\arabic{equation}}
\renewcommand{\thefigure}{S\arabic{figure}}
\renewcommand{\bibnumfmt}[1]{[S#1]}
\renewcommand{\citenumfont}[1]{S#1}

\author{A. Mendoza-Coto}%
\affiliation{Departamento de F\'\i sica, Universidade Federal de Santa Catarina, 88040-900 Florian\'opolis, Brazil}%
\affiliation{Max Planck Institute for the Physics of Complex Systems, Nothnitzerstr. 38, 01187 Dresden, Germany}

\author{M. Grossklags}%
\affiliation{Departamento de F\'\i sica, Universidade Federal de Santa Catarina, 88040-900 Florian\'opolis, Brazil}%

\author{J. Stefaniak}%
\affiliation{Institute for Quantum Electronics, Eidgenossische Technische Hochschule Zurich, Otto-Stern-Weg 1, CH-8093 Zurich, Switzerland}

\author{T. Donner}%
\affiliation{Institute for Quantum Electronics, Eidgenossische Technische Hochschule Zurich, Otto-Stern-Weg 1, CH-8093 Zurich, Switzerland}

\author{F. Piazza}%
\affiliation{Theoretical Physics III, Center for Electronic Correlations and Magnetism,
Institute of Physics, University of Augsburg, 86135 Augsburg, Germany}
\affiliation{Max Planck Institute for the Physics of Complex Systems, Nothnitzerstr. 38, 01187 Dresden, Germany}

\maketitle
\section{I. Nature of the quasicrystal-homogeneous phase transition}
\begin{figure}[!h]
    \centering
    \includegraphics[width=0.60\textwidth]{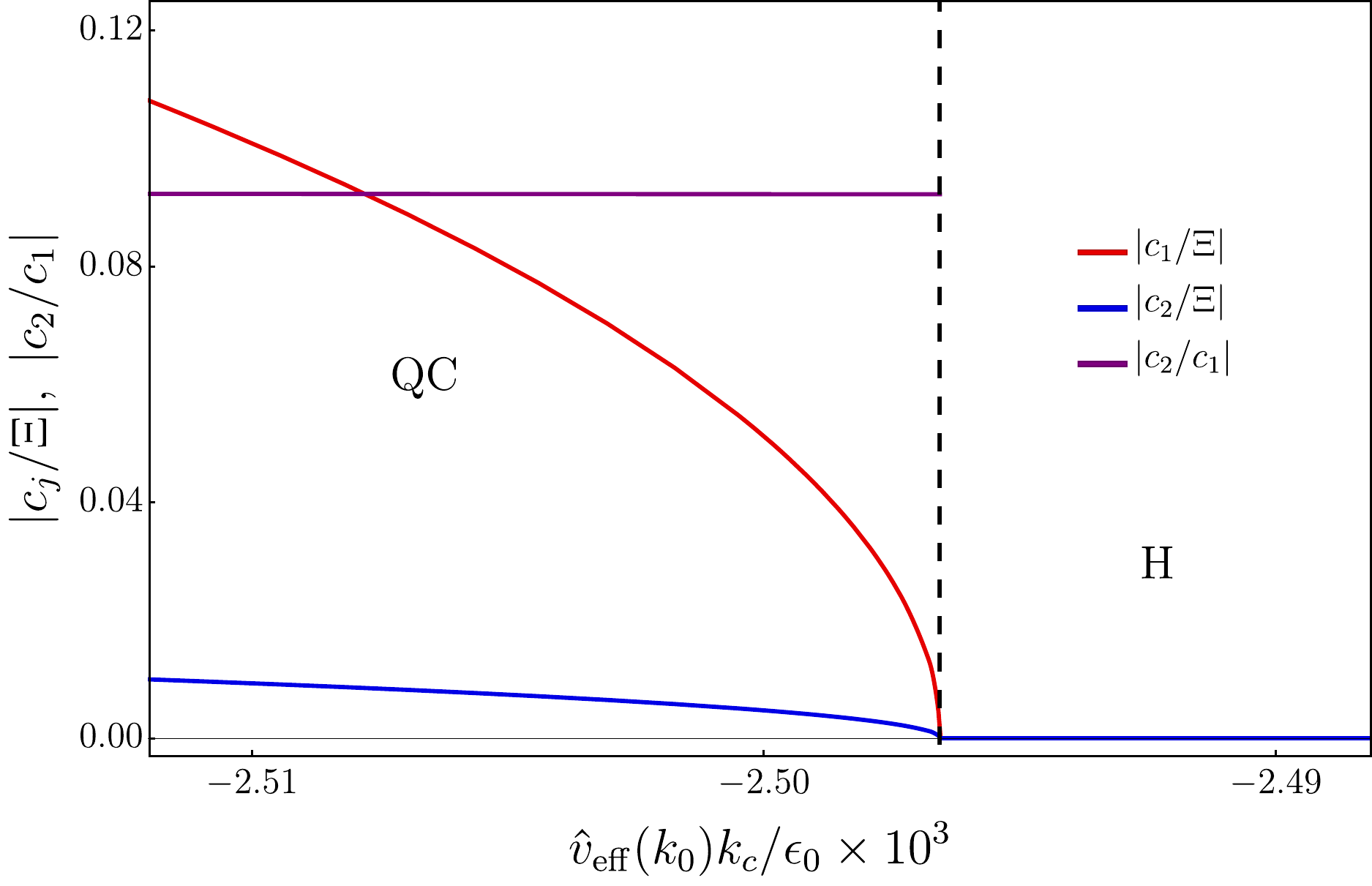}
    \caption{Behavior of the two main Fourier amplitudes of the density pattern varying $\hat{v}_{\mathrm{eff}}(k_0)k_c/\epsilon_0$ across the homogeneous to quasicrystal phase transition for $n/k_{c}=0.095\times10^{-3}$. The red and blue lines corresponds to the main and secondary Fourier amplitudes, while the purple line displays the ratio of the amplitudes as we approach the critical line.}
    \label{fig6}
\end{figure}

As discussed in the main text, the transition between the homogeneous and quasicrystal phase presented in Fig.~$1$(b) is second order. This means that the main Fourier amplitudes of the QC phase goes to zero continuously at the phase boundary. The corresponding behavior resulting from the numerical calculation of the ground-state wavefunction is shown in Fig.~\ref{fig6} as $\hat{v}_{\mathrm{eff}}(k_{0})k_{c}/\epsilon_{0}$ is varied at a fixed density $n/k_{c}=0.095\times10^{-3}$. As can be observed, both amplitudes obeys a mean-field scaling of the form $c_{j}\propto(\hat{v}_{\mathrm{eff}}(k_{0})\vert_{c}-\hat{v}_{\mathrm{eff}}(k_{0}))^{1/2}$, which means that their ratio is approximately constant in the critical region as shown by the purple curve in Fig.~\ref{fig6}. 

\section{II. Elastic theory}
In this section, we show the main steps followed in the construction of the low energy effective action of the $1$D model described by Eq.~$(1)$ of the main text. In dimensionless units, the grand canonical partition function of this model reads as $\mathcal{Z}=\int D\Psi D\Psi^{\ast}e^{-\mathcal{S}[\Psi,\Psi^{\ast}]}$, where the action is given by
\begin{equation}\label{general_model_action}
    \mathcal{S}[\Psi,\Psi^{\ast}]=\int\limits_{0}^{\infty}d\tau\int\limits dx~\Psi^{\ast}(x,\tau)\bigg(\partial_{\tau}-\frac{1}{2}\partial^{2}_{x}-\mu+\frac{1}{2}\int dx' \left(V_{\mathrm{eff}}(x,x')+g_{\mathrm{aa}}\delta(x-x')\right)\lvert\Psi(x',\tau)\rvert^{2}\bigg)\Psi(x,\tau),
\end{equation}
in which the dimensionless non-local two body interaction writes as $V_{\mathrm{eff}}(x,x')=\cos(2\pi x)\cos(2\pi x')v_{\mathrm{eff}}(x-x')$. It is not difficult to verify that the configuration $\Psi_{0}(x)$ minimizing the action $S[\Psi,\Psi^*]$, while subjected to the constraint $\int\limits dx\lvert\Psi\left(x,\tau\right)\rvert^{2}=N$, is given by $\Psi_{0}(x)=\sqrt{n}\psi_{0}(x)$, where $\psi_{0}(x)$ represents the ground-state wavefunction minimizing the energy functional in Eq.~$(1)$ of the main text and $n=N/L$, stand for the average linear density of particles of the system. As a consequence, the ground-state wavefunction $\Psi_{0}(x)$ in a quasicrystal configuration can be written as
\begin{equation}\label{unperturbed_wavefunction}
    \Psi_{0}(x)=\frac{\sqrt{n}}{2\Xi}\sum\limits_{\boldsymbol{\mathrm{n}}}c_{\boldsymbol{\mathrm{n}}}e^{ik_{0}G_{\boldsymbol{\mathrm{n}}}x},
\end{equation}
where $\boldsymbol{\mathrm{n}}=\{n_{1},n_{2}\}$ encloses the two indexes characterizing the wave vectors $G_{\boldsymbol{\mathrm{n}}}=n_{1}+\gamma n_{2}$ of the Fourier expansion and $\Xi=\sum_{\boldsymbol{\mathrm{n}}}c^{2}_{\boldsymbol{\mathrm{n}}}/4$ ensures the normalization condition for $\Psi_{0}(x)$. Finally, $k_{0}$ represents the characteristic wave vector of the modulated solution while $\gamma$ corresponds to a variational parameter characterizing the incommensurability of the two main wave vectors of the QC pattern, $k_{0}$ and $\gamma k_{0}$.

In analogy to recent developments in the elasticity theory of quantum crystals and quasicrystals~\cite{pretko2018,mendoza2025}, we propose the following ansatz for the wavefunction field upon inclusion of long wavelength fluctuations 
\begin{equation}\label{wavefunction_fluctuation_ansatz}
    \Psi(x,\tau)=\frac{\sqrt{n}}{2\Xi}\sum\limits_{\boldsymbol{\mathrm{n}}}c_{\boldsymbol{\mathrm{n}}}\left(1+\frac{\delta n(x,\tau)}{n}+\boldsymbol{\mathrm{Q}}_{\boldsymbol{\mathrm{n}}}\cdot\boldsymbol{\mathrm{Z}}^{-1}\cdot\boldsymbol{\mathrm{\Pi}}(x,\tau)\right)^{1/2}\exp\bigg(i\boldsymbol{\mathrm{Q}}_{\boldsymbol{\mathrm{n}}}\cdot\big(\boldsymbol{\mathrm{X}}+\boldsymbol{\mathrm{U}}(x,\tau)\big)+i\theta(x,\tau)\bigg),
\end{equation}
where
\begin{equation}\label{notation_definition}
    \boldsymbol{\mathrm{Q}}_{\boldsymbol{\mathrm{n}}}=\boldsymbol{\mathrm{k}}^{\mathrm{T}}_{\boldsymbol{\mathrm{n}}}\cdot\boldsymbol{\mathrm{T}},\hspace{0.2cm}
    \boldsymbol{\mathrm{k}}_{\boldsymbol{\mathrm{n}}}\equiv k_{0}
    \begin{pmatrix}
        n_{1}\\
        n_{2}
    \end{pmatrix},\hspace{0.2cm}
    \boldsymbol{\mathrm{T}}\equiv
    \begin{pmatrix}
        1 & -\gamma\\
        \gamma & 1
    \end{pmatrix}, \hspace{0.2cm}
    \boldsymbol{\mathrm{X}}\equiv
    \begin{pmatrix}
        x\\
        0
    \end{pmatrix},\hspace{0.2cm}
    \boldsymbol{\mathrm{U}}(x,\tau)\equiv
    \begin{pmatrix}
        u(x,\tau)\\
        w(x,\tau)
    \end{pmatrix}\hspace{0.2cm}\text{and}\hspace{0.3cm}
    \boldsymbol{\mathrm{\Pi}}\equiv
    \begin{pmatrix}
       \Pi(x,\tau)\\
        \Pi_\perp(x,\tau)
    \end{pmatrix}.
\end{equation}
In Eq.~\eqref{wavefunction_fluctuation_ansatz}, fluctuations in the phases of the quasicrystal modulation are captured by the two dimensional deformation field $\boldsymbol{\mathrm{U}}(x,\tau)$, whose first and second components corresponds to the phonon and phason fields, respectively, while the field $\theta(x,\tau)$ stands for the condensate phase fluctuation. Moreover, each complex exponential term containing a linear combination of the phase fluctuations is coupled to a square root contribution including the corresponding conjugate density fluctuation to that specific exponential. Such structure generalize the well known ansatz for the wavefunction fields of a homogeneous condensate in terms of the density and phase fluctuations~\cite{stoof2009}. The rank-$2$ matrix $\boldsymbol{\mathrm{Z}}$, whose inverse enters the wavefunction ansatz, has components chosen to ensure that the density and phase fields, $\boldsymbol{\mathrm{\Pi}}(x,\tau)$ and $\boldsymbol{\mathrm{U}}(x,\tau)$, are canonically conjugate. As a result, it generates a dynamical term in the long-wavelength effective action with the appropriate structure.

In order to investigate the low energy behavior of the system, we work in the hydrodynamic regime in which the fluctuation fields vary slowly in space and imaginary time. Mathematically, this allow us in the calculation of the effective action to neglect all quadratic contributions in the fluctuation fields that are multiplied by a zero mean spatial oscillatory function. Next, we proceed with the evaluation of each contribution to the action in terms of the fluctuation fields. The first term in Eq.~\eqref{general_model_action} can be rewritten in the symmetrized form
\begin{equation}\label{action1_expression1}
    \delta\mathcal{S}_{1}=\frac{1}{2}\int d\tau\int dx\bigg(\Psi^{\ast}(x,\tau)\partial_{\tau}\Psi(x,\tau)-\Psi(x,\tau)\partial_{\tau}\Psi^{\ast}(x,\tau)\bigg),
\end{equation}
which up to second order in the fluctuation fields reads
\begin{equation}\label{action1_expression2}
    \delta\mathcal{S}_{1}=\int d\tau\int dx~in\bigg[\frac{\delta n(x,\tau)\partial_{\tau}\theta(x,\tau)}{n}+\sum_{\boldsymbol{\mathrm{n}}}\frac{c^{2}_{\boldsymbol{\mathrm{n}}}}{4\Xi^2}\big(\boldsymbol{\mathrm{Q}}_{\boldsymbol{\mathrm{n}}}\cdot\boldsymbol{\mathrm{Z}}^{-1}\cdot\boldsymbol{\mathrm{\Pi}}(x,\tau)\big)\big(\boldsymbol{\mathrm{Q}}_{\boldsymbol{\mathrm{n}}}\cdot\partial_{\tau}\boldsymbol{\mathrm{U}}(x,\tau)\big)\bigg].
\end{equation}
Simplifying the above expression, it is possible to conclude that when the elements of $\boldsymbol{\mathrm{Z}}$ has the form
\begin{equation}
    \begin{split}
        Z_{11}=&\sum\limits_{n_{1},n_{2}}\frac{c^{2}_{n_{1}n_{2}}}{4\Xi^2}\left(n_{1}+\gamma n_{2}\right)^{2}k^{2}_{0},\\
        Z_{22}=&\sum\limits_{n_{1},n_{2}}\frac{c^{2}_{n_{1}n_{2}}}{4\Xi^2}\left(-\gamma n_{1}+ n_{2}\right)^{2}k^{2}_{0},\\
        Z_{12}=&\sum\limits_{n_{1},n_{2}}\frac{c^{2}_{n_{1}n_{2}}}{4\Xi^2}\left(n_{1}+\gamma n_{2}\right)\left(-\gamma n_{1}+n_{2}\right)k^{2}_{0}=Z_{21},
    \end{split}
\end{equation}
the dynamical term in the action reduces to
\begin{equation}\label{action1_expression3}
    \delta\mathcal{S}_{1}=\int d\tau\int dx~\bigg(i\delta n(x,\tau)\partial_{\tau}\theta(x,\tau)+i\boldsymbol{\mathrm{\Pi}}(x,\tau)\cdot\partial_{\tau}\boldsymbol{\mathrm{U}}(x,\tau)\bigg),
\end{equation}
which as desired, has the expected structure in terms of imaginary time phase derivatives and conjugate density fluctuations. The second contribution in Eq.~\eqref{general_model_action} can be written in the symmetrical form  
\begin{equation}
    \delta\mathcal{S}_{2}=\frac{1}{2}\int d\tau\int dx\bigg(\partial_{x}\Psi^{\ast}(x,\tau)\partial_{x}\Psi(x,\tau)\bigg),
\end{equation}
whose expansion up to second order in the fluctuation fields yields
\begin{equation}
    \begin{split}
        \delta\mathcal{S}_{2}=\int d\tau\int dx~\bigg[&\frac{1}{8n}\big(\partial_{x}\delta n(x,\tau)\big)^{2}+\frac{1}{8}\partial_{x}\boldsymbol{\mathrm{\Pi}}(x,\tau)\cdot\boldsymbol{\mathrm{Z}}^{-1}\cdot\partial_{x}\boldsymbol{\mathrm{\Pi}}(x,\tau)+\frac{n}{2}\big(\partial_{x}\theta(x,\tau)\big)^{2}+\frac{1}{2}\partial_{x}\boldsymbol{\mathrm{U}}(x,\tau)\cdot\boldsymbol{\mathrm{Z}}\cdot\partial_{x}\boldsymbol{\mathrm{U}}(x,\tau)\\
        &+\frac{\delta n(x,\tau)}{n}\boldsymbol{\mathrm{e}}_{1}\cdot\boldsymbol{\mathrm{Z}}\cdot\partial_x\boldsymbol{\mathrm{U}}(x,\tau)+\boldsymbol{\mathrm{e}}_{1}\cdot\boldsymbol{\mathrm{\Pi}}(x,\tau)\partial_{x}\theta(x,\tau)\bigg],
    \end{split}
\end{equation}
where $\boldsymbol{\mathrm{e}}_{1}$ stands for the unit vector $(1,0)$. The third term in our action, $\delta\mathcal{S}_{3}=-\mu\int d\tau\int dx~\lvert\Psi(x,\tau)\rvert^{2}$, does not produce any quadratic contribution in the fluctuation fields. Finally, we need to address the contributions produced by the two body interaction in Eq.~\eqref{general_model_action}. In this case, we should consider separately the non-local and the contact interaction terms. Therefore, we define
\begin{equation}\label{action4_expression1}
    \begin{split}
        \delta\mathcal{S}_{4,1}&=\frac{1}{2}\int d\tau\int dxdx'V_{\mathrm{eff}}(x,x')\lvert\Psi(x',\tau)\rvert^{2}\lvert\Psi(x,\tau)\rvert^{2},\\
        \delta\mathcal{S}_{4,2}&=\frac{1}{2}\int d\tau\int dxdx'g_{\mathrm{aa}}\lvert\Psi(x',\tau)\rvert^{2}\lvert\Psi(x,\tau)\rvert^{2},
    \end{split}
\end{equation}
and proceed with the quadratic expansion in the fluctuation fields of $\delta\mathcal{S}_{4,1}$. After a lengthy calculation, we find
\begin{equation}
    \begin{split}
        \delta\mathcal{S}_{4,1}=&\int d\tau\int\frac{dq}{2\pi}\frac{1}{2}\bigg[G_{00}\delta\hat{n}(q,\tau)\delta\hat{n}(-q,\tau)+2iq\beta_{1}u(q,\tau)\delta\hat{n}(-q,\tau)+2iq\beta_{2}w(q,\tau)\delta\hat{n}(-q,\tau)+\kappa_{1}\hat{\delta}_{1}(q,\tau)^{2}\\
        &+\Gamma_{11}q^{2}\hat{\delta}_{1}(q,\tau)\hat{\delta}_{1}(-q,\tau)+\Gamma_{22}q^{2}\hat{\delta}_{2}(q,\tau)\hat{\delta}_{2}(-q,\tau)+2\Gamma_{12}q^{2}\hat{\delta}_{1}(q,\tau)\hat{\delta}_{2}(-q,\tau)+G_{11}\hat{\Pi}(q,\tau)\hat{\Pi}(-q,\tau)\\
        &+G_{22}\hat{\Pi}_{\perp}(q,\tau)\hat{\Pi}_{\perp}(-q,\tau)+2G_{12}\hat{\Pi}(q,\tau)\hat{\Pi}_{\perp}(-q,\tau)\bigg],
    \end{split}
\end{equation}
where the couplings in the above expression are given as
\begin{equation}
    \begin{split}
        G_{00}=&\sum_{\boldsymbol{\mathrm{n}}_{1},\boldsymbol{\mathrm{n}}_{2},\boldsymbol{\mathrm{n}}_{3},\boldsymbol{\mathrm{n}}_{4},j_{1},j_{2}}\frac{c_{\boldsymbol{\mathrm{n}}_{1}}c_{\boldsymbol{\mathrm{n}}_{2}}c_{\boldsymbol{\mathrm{n}}_{3}}c_{\boldsymbol{\mathrm{n}}_{4}}}{64}\frac{1}{\Xi^{4}}\hat{v}_{\mathrm{eff}}(\boldsymbol{\mathrm{Q}}^{(1)}_{\boldsymbol{\mathrm{n}}_{1}}-\boldsymbol{\mathrm{Q}}^{(1)}_{\boldsymbol{\mathrm{n}}_{2}}-k_{0}j_{1})\delta(\boldsymbol{\mathrm{n}}_{1}-\boldsymbol{\mathrm{n}}_{2}+\boldsymbol{\mathrm{n}}_{3}-\boldsymbol{\mathrm{n}}_{4}-\boldsymbol{\mathrm{j}}_{1}-\boldsymbol{\mathrm{j}}_{2},\boldsymbol{\mathrm{0}}),\\
        \beta_{1}=&\sum_{\boldsymbol{\mathrm{n}}_{1},\boldsymbol{\mathrm{n}}_{2},\boldsymbol{\mathrm{n}}_{3},\boldsymbol{\mathrm{n}}_{4},j_{1},j_{2}}\frac{c_{\boldsymbol{\mathrm{n}}_{1}}c_{\boldsymbol{\mathrm{n}}_{2}}c_{\boldsymbol{\mathrm{n}}_{3}}c_{\boldsymbol{\mathrm{n}}_{4}}}{64}\frac{n}{\Xi^{4}}(\boldsymbol{\mathrm{Q}}^{(1)}_{\boldsymbol{\mathrm{n}}_{4}}-\boldsymbol{\mathrm{Q}}^{(1)}_{\boldsymbol{\mathrm{n}}_{3}})\partial_{q}\hat{v}_{\mathrm{eff}}(\boldsymbol{\mathrm{Q}}^{(1)}_{\boldsymbol{\mathrm{n}}_{1}}-\boldsymbol{\mathrm{Q}}^{(1)}_{\boldsymbol{\mathrm{n}}_{2}}-k_{0}j_{1})\delta(\boldsymbol{\mathrm{n}}_{1}-\boldsymbol{\mathrm{n}}_{2}+\boldsymbol{\mathrm{n}}_{3}-\boldsymbol{\mathrm{n}}_{4}-\boldsymbol{\mathrm{j}}_{1}-\boldsymbol{\mathrm{j}}_{2},\boldsymbol{\mathrm{0}}),\\
        \beta_{2}=&\sum_{\boldsymbol{\mathrm{n}}_{1},\boldsymbol{\mathrm{n}}_{2},\boldsymbol{\mathrm{n}}_{3},\boldsymbol{\mathrm{n}}_{4},j_{1},j_{2}}\frac{c_{\boldsymbol{\mathrm{n}}_{1}}c_{\boldsymbol{\mathrm{n}}_{2}}c_{\boldsymbol{\mathrm{n}}_{3}}c_{\boldsymbol{\mathrm{n}}_{4}}}{64}\frac{n}{\Xi^{4}}(\boldsymbol{\mathrm{Q}}^{(2)}_{\boldsymbol{\mathrm{n}}_{4}}-\boldsymbol{\mathrm{Q}}^{(2)}_{\boldsymbol{\mathrm{n}}_{3}})\partial_{q}\hat{v}_{\mathrm{eff}}(\boldsymbol{\mathrm{Q}}^{(1)}_{\boldsymbol{\mathrm{n}}_{1}}-\boldsymbol{\mathrm{Q}}^{(1)}_{\boldsymbol{\mathrm{n}}_{2}}-k_{0}j_{1})\delta(\boldsymbol{\mathrm{n}}_{1}-\boldsymbol{\mathrm{n}}_{2}+\boldsymbol{\mathrm{n}}_{3}-\boldsymbol{\mathrm{n}}_{4}-\boldsymbol{\mathrm{j}}_{1}-\boldsymbol{\mathrm{j}}_{2},\boldsymbol{\mathrm{0}}),\\
        \kappa_{1}=&\sum_{\boldsymbol{\mathrm{n}}_{1},\boldsymbol{\mathrm{n}}_{2},\boldsymbol{\mathrm{n}}_{3},\boldsymbol{\mathrm{n}}_{4},j_{1},j_{2}}\frac{c_{\boldsymbol{\mathrm{n}}_{1}}c_{\boldsymbol{\mathrm{n}}_{2}}c_{\boldsymbol{\mathrm{n}}_{3}}c_{\boldsymbol{\mathrm{n}}_{4}}}{64}\frac{n^{2}}{\Xi^{4}}k^{2}_{0}(\boldsymbol{\mathrm{n}}^{(1)}_{1}-\boldsymbol{\mathrm{n}}^{(1)}_{2})(\boldsymbol{\mathrm{n}}^{(1)}_{4}-\boldsymbol{\mathrm{n}}^{(1)}_{3})\hat{v}_{\mathrm{eff}}(\boldsymbol{\mathrm{Q}}^{(1)}_{\boldsymbol{\mathrm{n}}_{1}}-\boldsymbol{\mathrm{Q}}^{(1)}_{\boldsymbol{\mathrm{n}}_{2}}-k_{0}j_{1})\\
        &\times\delta(\boldsymbol{\mathrm{n}}_{1}-\boldsymbol{\mathrm{n}}_{2}+\boldsymbol{\mathrm{n}}_{3}-\boldsymbol{\mathrm{n}}_{4}-\boldsymbol{\mathrm{j}}_{1}-\boldsymbol{\mathrm{j}}_{2},\boldsymbol{\mathrm{0}}),\\
        \Gamma_{ij}=&\sum_{\boldsymbol{\mathrm{n}}_{1},\boldsymbol{\mathrm{n}}_{2},\boldsymbol{\mathrm{n}}_{3},\boldsymbol{\mathrm{n}}_{4},j_{1},j_{2}}\frac{c_{\boldsymbol{\mathrm{n}}_{1}}c_{\boldsymbol{\mathrm{n}}_{2}}c_{\boldsymbol{\mathrm{n}}_{3}}c_{\boldsymbol{\mathrm{n}}_{4}}}{128}\frac{n^{2}}{\Xi^{4}}k^{2}_{0}(\boldsymbol{\mathrm{n}}^{(i)}_{1}-\boldsymbol{\mathrm{n}}^{(i)}_{2})(\boldsymbol{\mathrm{n}}^{(j)}_{3}-\boldsymbol{\mathrm{n}}^{(j)}_{4})\epsilon_{i}\epsilon_{j}\partial^{2}_{q}\hat{v}_{\mathrm{eff}}(\boldsymbol{\mathrm{Q}}^{(1)}_{\boldsymbol{\mathrm{n}}_{1}}-\boldsymbol{\mathrm{Q}}^{(1)}_{\boldsymbol{\mathrm{n}}_{2}}-k_{0}j_{1})\\
        &\times\delta(\boldsymbol{\mathrm{n}}_{1}-\boldsymbol{\mathrm{n}}_{2}+\boldsymbol{\mathrm{n}}_{3}-\boldsymbol{\mathrm{n}}_{4}-\boldsymbol{\mathrm{j}}_{1}-\boldsymbol{\mathrm{j}}_{2},\boldsymbol{\mathrm{0}}),\\
        G_{ij}=&-\sum_{\boldsymbol{\mathrm{n}}_{1},\boldsymbol{\mathrm{n}}_{2},\boldsymbol{\mathrm{n}}_{3},\boldsymbol{\mathrm{n}}_{4},j_{1},j_{2}}\frac{c_{\boldsymbol{\mathrm{n}}_{1}}c_{\boldsymbol{\mathrm{n}}_{2}}c_{\boldsymbol{\mathrm{n}}_{3}}c_{\boldsymbol{\mathrm{n}}_{4}}}{256}\frac{n^{2}}{\Xi^{4}}k^{2}_{0}\big((\boldsymbol{\mathrm{Q}}_{\boldsymbol{\mathrm{n}}_{1}}\cdot\boldsymbol{\mathrm{Z}}^{-1})^{(i)}-(\boldsymbol{\mathrm{Q}}_{\boldsymbol{\mathrm{n}}_{2}}\cdot\boldsymbol{\mathrm{Z}}^{-1})^{(i)}\big)\big((\boldsymbol{\mathrm{Q}}_{\boldsymbol{\mathrm{n}}_{3}}\cdot\boldsymbol{\mathrm{Z}}^{-1})^{(j)}-(\boldsymbol{\mathrm{Q}}_{\boldsymbol{\mathrm{n}}_{4}}\cdot\boldsymbol{\mathrm{Z}}^{-1})^{(j)}\big)\\
        &\times\hat{v}_{\mathrm{eff}}(\boldsymbol{\mathrm{Q}}^{(1)}_{\boldsymbol{\mathrm{n}}_{1}}-\boldsymbol{\mathrm{Q}}^{(1)}_{\boldsymbol{\mathrm{n}}_{2}}-k_{0}j_{1})\delta(\boldsymbol{\mathrm{n}}_{1}-\boldsymbol{\mathrm{n}}_{2}+\boldsymbol{\mathrm{n}}_{3}-\boldsymbol{\mathrm{n}}_{4}-\boldsymbol{\mathrm{j}}_{1}-\boldsymbol{\mathrm{j}}_{2},\boldsymbol{\mathrm{0}}).
    \end{split}
\end{equation}
Here, the superscript index refers to the component of the corresponding two-dimensional vector, whereas the symbol $\epsilon_{i}$ is defined as $\epsilon_{i}=\delta_{i,1}+\gamma\delta_{i,2}$. Moreover, $\boldsymbol{\mathrm{j}}_{i}=j_{i}(1,0)$, where $j_{i}$ takes values $-1$ or $1$. In a similar way, the power expansion in the fluctuation fields for $\delta S_{4,2}$ lead us to
\begin{equation}
    \begin{split}
        \delta\mathcal{S}_{4,2}=&\int d\tau\int\frac{dq}{2\pi}\frac{1}{2}\bigg[\tilde{G}_{00}\delta\hat{n}(q,\tau)\delta\hat{n}(-q,\tau)+\tilde{G}_{11}\hat{\Pi}(q,\tau)\hat{\Pi}(-q,\tau)+\tilde{G}_{22}\hat{\Pi}_{\perp}(q,\tau)\hat{\Pi}_{\perp}(-q,\tau)\\
        &+2\tilde{G}_{12}\hat{\Pi}(q,\tau)\hat{\Pi}_{\perp}(-q,\tau)\bigg],
    \end{split}
\end{equation}
where
\begin{equation}
    \begin{split}
        \tilde{G}_{00}=&\sum_{\boldsymbol{\mathrm{n}}_{1},\boldsymbol{\mathrm{n}}_{2},\boldsymbol{\mathrm{n}}_{3},\boldsymbol{\mathrm{n}}_{4}}\frac{c_{\boldsymbol{\mathrm{n}}_{1}}c_{\boldsymbol{\mathrm{n}}_{2}}c_{\boldsymbol{\mathrm{n}}_{3}}c_{\boldsymbol{\mathrm{n}}_{4}}}{16}\frac{1}{\Xi^{4}}\hat{v}_{\mathrm{eff}}(\boldsymbol{\mathrm{Q}}^{(1)}_{\boldsymbol{\mathrm{n}}_{1}}-\boldsymbol{\mathrm{Q}}^{(1)}_{\boldsymbol{\mathrm{n}}_{2}})\delta(\boldsymbol{\mathrm{n}}_{1}-\boldsymbol{\mathrm{n}}_{2}+\boldsymbol{\mathrm{n}}_{3}-\boldsymbol{\mathrm{n}}_{4},\boldsymbol{\mathrm{0}}),\\
        \tilde{G}_{ij}=&-\sum_{\boldsymbol{\mathrm{n}}_{1},\boldsymbol{\mathrm{n}}_{2},\boldsymbol{\mathrm{n}}_{3},\boldsymbol{\mathrm{n}}_{4}}\frac{c_{\boldsymbol{\mathrm{n}}_{1}}c_{\boldsymbol{\mathrm{n}}_{2}}c_{\boldsymbol{\mathrm{n}}_{3}}c_{\boldsymbol{\mathrm{n}}_{4}}}{16}\frac{n^{2}}{\Xi^{4}}k^{2}_{0}\big((\boldsymbol{\mathrm{Q}}_{\boldsymbol{\mathrm{n}}_{1}}\cdot\boldsymbol{\mathrm{Z}}^{-1})^{(i)}-(\boldsymbol{\mathrm{Q}}_{\boldsymbol{\mathrm{n}}_{2}}\cdot\boldsymbol{\mathrm{Z}}^{-1})^{(i)}\big)\big((\boldsymbol{\mathrm{Q}}_{\boldsymbol{\mathrm{n}}_{3}}\cdot\boldsymbol{\mathrm{Z}}^{-1})^{(j)}-(\boldsymbol{\mathrm{Q}}_{\boldsymbol{\mathrm{n}}_{4}}\cdot\boldsymbol{\mathrm{Z}}^{-1})^{(j)}\big)\\
        &\times\hat{v}_{\mathrm{eff}}(\boldsymbol{\mathrm{Q}}^{(1)}_{\boldsymbol{\mathrm{n}}_{1}}-\boldsymbol{\mathrm{Q}}^{(1)}_{\boldsymbol{\mathrm{n}}_{2}})\delta(\boldsymbol{\mathrm{n}}_{1}-\boldsymbol{\mathrm{n}}_{2}+\boldsymbol{\mathrm{n}}_{3}-\boldsymbol{\mathrm{n}}_{4},\boldsymbol{\mathrm{0}}).
    \end{split}
\end{equation}
In this way, the effective quadratic action can be computed as $\delta\mathcal{S}=\delta\mathcal{S}_{1}+\delta\mathcal{S}_{2}+\delta\mathcal{S}_{4,1}+\delta\mathcal{S}_{4,2}$, which to leading order in the fluctuation fields reads
\begin{equation}
    \begin{split}
        \delta\mathcal{S}=&\int d\tau dx\bigg[i\delta n(x,\tau)\partial_{\tau}\theta(x,\tau)+i\Pi(x,\tau)\partial_{\tau}u(x,\tau)+i\Pi_{\perp}(x,\tau)\partial_{\tau}w(x,\tau)+\frac{1}{2}\bigg(n\big(\partial_{x}\theta(x,\tau)\big)^{2}+g_{00}\big(\delta n(x,\tau)\big)^{2}\\
        &+g_{11}\Pi(x,\tau)^{2}+g_{22}\Pi_{\perp}(x,\tau)^{2}+2g_{12}\Pi(x,\tau)\Pi_{\perp}(x,\tau)+2\Pi(x,\tau)\partial_{x}\theta(x,\tau)+2\alpha_{1}\delta n(x,\tau)\partial_{x}u(x,\tau)+\kappa_{1}\delta_{1}(x,\tau)^{2}\\
        &+2\alpha_{2}\delta n(x,\tau)\partial_{x}w(x,\tau)+\kappa_{2}\big(\partial_{x}\delta_{2}(x,\tau)\big)^{2}\bigg)\bigg],
    \end{split}
\end{equation}
where $g_{ij}=G_{ij}+\tilde{G}_{ij}$, $\alpha_{1}=Z_{11}+\beta_{1}$, $\alpha_{2}=Z_{12}+\beta_{2}$ and $\kappa_{2}=\Gamma_{22}+\gamma^{2}(\gamma(\gamma Z_{11}+2Z_{12})+Z_{22})/(1+\gamma^{2})^{2}$. Finally, it is worth mentioning that the form of the action obtained here is consistent with those derived for $1$D supersolids after integrating out the field $\Pi(x,\tau)$~\cite{yoo2010,platt2024}. In general terms, the expressions for the couplings of our effective theory presented above assume that the ground-state Fourier amplitudes do not change when a sufficiently small perturbation is applied. This condition cannot, in general, be guaranteed. However, this does not alter the main conclusion regarding the structure of the effective action, since it is a direct consequence of the symmetry properties of the system~\cite{yoo2010}. More generally, the coupling parameters can instead be determined by exploiting the connection between the response of the ground-state energy and the response of the action to specific perturbations; see Ref.~\cite{platt2024} for details about this procedure.

\section{III. Energy-momentum dispersion relations}
As discussed in the main text, the quasicrystal phase exhibits three distinct low-energy excitation modes. The gapped branch has the dispersion relation $\omega_{1}(q)=\omega_{1}(0)+cq^{2}$, with $\omega_{1}(0)=\sqrt{g_{11}-2\Gamma g_{12}+\Gamma^{2}g_{22}}$. The remaining two are gapless and display linear behavior, $\omega_{\pm}(q)=c_{\pm}q$, where the sound velocities satisfy $c^{2}_{\pm}=(A\pm\sqrt{A^{2}-4B})/2$. The coefficients $A$ and $B$ are rational functions of the couplings of the effective low-energy theory, given by
\begin{equation}
        \begin{split}
            A&=\frac{\Gamma^{2}\big(2(\alpha_{2}+\alpha_{1}\Gamma)(-g_{12}+\Gamma g_{22})-g_{00}\big) -(1+\Gamma^{2})^{2}(g_{12}^{2}-g_{11}g_{22})\kappa_{2}+\Gamma^{2}\big(g_{11}-2\Gamma g_{12}+\Gamma^{2} g_{22}\big)g_{00}n}{\Gamma^{2}\big(g_{11}-2\Gamma g_{12}+\Gamma^{2} g_{22}\big)},\\ 
            B&=\frac{\big(-\Gamma^{2}(\alpha_{2}+\alpha_{1}\Gamma)^{2}+(1+\Gamma^{2})^{2}g_{00}\kappa_{2}\big)\big(-g^{2}_{12}n+g_{22}(-1+g_{11}n)\big)}{\Gamma^{2}\big(g_{11}-2\Gamma g_{12}+\Gamma^{2} g_{22}\big)}.
        \end{split}
\end{equation}

\section{IV. Density profile under phonon and phason propagation}
In this section, we discuss the space-time behavior of the density profile associated to the propagation of a pure phonon and phason excitation. For demonstration purposes, we consider a simplified quasicrystal ansatz whose Fourier expansion displays only two Fourier modes with the wave vectors $k_{1}$ and $k_{2}$. In the case of a phonon propagation, we consider $u(x,t)=a\cos(\omega t-kx)$ and $w(x,t)=0$, which yields a density profile evolution of the form
\begin{equation}
    \rho_{\mathrm{phonon}}(x,t)=\sum\limits_{\boldsymbol{\mathrm{n}}}b_{\boldsymbol{\mathrm{n}}}\cos\big(k_{\boldsymbol{\mathrm{n}}}(x+u(x,t))\big)\big(1+\partial_{x}u(x,t)\big),
\end{equation}
once we consider that the envelope of the density profile re-accommodates to guarantee the local conservation of the number of particles. Moreover, to study the behavior of the density profile under the propagation of a phasonic excitation we consider the scenario in which $u(x,t)=0$ and $w(x,t)=a\cos(\omega t-kx)$. In this case, under the same constraints considered previously, it is straightforward to conclude that the density profile writes as
\begin{equation}
    \rho_{\mathrm{phason}}(x,t)=\sum\limits_{\boldsymbol{\mathrm{n}}}b_{\boldsymbol{\mathrm{n}}}\cos\big(k_{\boldsymbol{\mathrm{n}}}x+k_{\perp,\boldsymbol{\mathrm{n}}} w(x,t)\big)\bigg(1+\frac{k_{\perp,\boldsymbol{\mathrm{n}}}}{k_{\boldsymbol{\mathrm{n}}}}\partial_{x}w(x,t)\bigg),
\end{equation}
where, as in the rest of manuscript, $k_{\boldsymbol{\mathrm{n}}}=n_{1}k_{1}+n_{2}k_{2}$ and $k_{\perp,\boldsymbol{\mathrm{n}}}=-n_{1}k_{2}+n_{2}k_{1}$.

\begin{figure}[!h]
    \centering
    \includegraphics[width=1.00\textwidth]{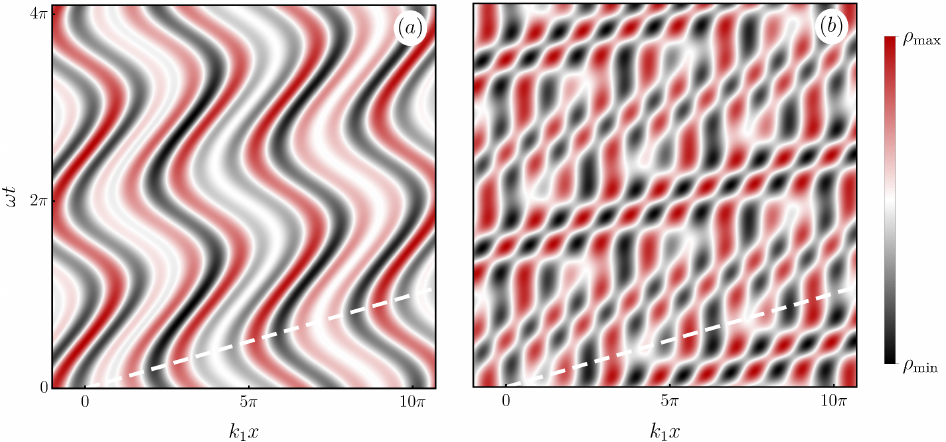}
    \caption{Space-time local density behavior of a $1$D QC phase under the propagation of a long-wavelength excitation for a pure (b) phonon and (b) phason excitation. The white dashed line corresponds to the curve $x=c_{s}t$, where $c_{s}$ represents the sound velocity of the excitation mode. The parameters used for excitation fields $u(x,t)$ or $w(x,t)$, frequency and velocity with respect the one of $\delta_{2}(x,t)$ are the same as those of Fig.~$3$ for gapless excitations in the main text.}
    \label{fig5}
\end{figure}

The density profile under the propagation of a pure phononic and phasonic excitations is displayed in Fig.~\ref{fig5}(a) and Fig.~\ref{fig5}(b), respectively. For calculations, we have employed the following parameters $a=3$, $k=0.1$, $k_{1}=1$, $k_{2}=\gamma=(1+\sqrt{5})/2$, as well as a sound velocity for the excitations $c_{s}=5$ and frequency $\omega=c_{s}k$. The parameters employed here are the same considered in the construction of Fig.~$3$ of the main text.

\section{V. $\delta_{i}-\delta_{i}$ correlation behavior}
As discussed in the main text, the structure of the low-energy effective action in Eq.~$(6)$ implies that the correlations $\langle\hat{\delta_{i}}(q,\omega)\hat{\delta_{i}}(q',\omega')\rangle$ are diagonal in momentum and frequency space. Therefore, they can be written as
\begin{equation}
    \langle\hat{\delta}_{i}(q,\omega)\hat{\delta}_{i}(q',\omega')\rangle
    =(2\pi)^{2}\delta(q+q')\delta(\omega+\omega')C_{i}(q,\omega),
\end{equation}
where in the long-wavelength limit $q\rightarrow0$, the correlation functions take the form
\begin{equation}
    C_{1}(q,\omega)=\left(\kappa_{1}+\gamma_{11}\omega^{2}+\gamma_{12}q^{2}\right)^{-1},\hspace{1.0cm}
    C_{2}(q,\omega)=\left(\gamma_{21}\omega^{2}+\gamma_{22}q^{2}\right)^{-1}.
\end{equation}
The coefficients $\gamma_{ij}$ encoding the leading spatial and dynamical behavior of the correlation functions are given by 
\begin{equation}
    \begin{split}
        \gamma_{11}=&\frac{1}{g_{11}+\Gamma\left(-2g_{12}+\Gamma g_{22}\right)},\\
        \gamma_{21}=&\frac{\Gamma^{2}\left(g_{11}-2\Gamma g_{12}+\Gamma^{2}g_{22}\right)\left(-g^{2}_{12}+g_{11}g_{22}\right)}{\left(1+\Gamma^{2}\right)^{2}\left(g^{2}_{12}-g_{11}g_{22}\right)^{2}},
    \end{split}
\end{equation}
and
\begin{equation}
    \begin{split}
        \gamma_{12}=&\frac{1}{\Gamma^{2}\left(g_{11}+\Gamma\left(-2g_{12}+\Gamma g_{22}\right)\right)^{2}}\bigg[2\alpha_{1}\Gamma^{2}\left(g_{11}-\Gamma g_{12}\right)+\Gamma\big(2\alpha_{2}\Gamma(g_{12}-\Gamma g_{22})+\Gamma g_{00}\big)\\
        &+\kappa_{2}\left(g_{12}-\Gamma\left(-g_{11}+\Gamma g_{12}+g_{22}\right)\right)^{2}\bigg],\\
        \gamma_{22}=&\frac{1}{\left(1+\Gamma^{2}\right)^{2}\left(g^{2}_{12}-g_{11}g_{22}\right)^{2}}\bigg[\Gamma^{2}\left(-g_{12}+\Gamma g_{22}\right)\left(-2\left(\Gamma\alpha_{1}+\alpha_{2}\right)\left(g^{2}_{12}-g_{11}g_{22}\right)+g_{00}\left(-g_{12}+\Gamma g_{22}\right)\right)+\kappa_{2}\left(1+\Gamma^{2}\right)^{2}\\
        &\times\left(g^{2}_{12}-g_{11}g_{22}\right)\bigg],
    \end{split}
\end{equation}
where the coefficients $g_{ij}$ and $\kappa_{j}$ were determined in Sec.~II of the Supplemental Material, and $\Gamma=k_{2}/k_{1}$.


\end{document}